\documentclass[a4paper,12pt,twoside]{article} 
\usepackage{a4wide}
\usepackage{graphicx}
\usepackage{rotating}
\newcommand{\beq}{\begin{equation}}
\newcommand{\eeq}{\end{equation}}
\def\gs{\mathrel{ \rlap{\raise
0.511ex \hbox{$>$}}{\lower 0.511ex \hbox{$\sim$}}}} \def\ls{\mathrel{
\rlap{\raise 0.511ex \hbox{$<$}}{\lower 0.511ex \hbox{$\sim$}}}}

\newcommand{\ba}{\begin{array}{c}}
\newcommand{\baz}{\begin{array}{cc}}
\newcommand{\bad}{\begin{array}{ccc}}
\newcommand{\bea}{\begin{equation} \begin{array}{c}}
\newcommand{\eea}{ \end{array} \end{equation}}
\newcommand{\ea}{\end{array}}

\def\gtap{\mathrel{ \rlap{\raise 0.511ex \hbox{$>$}}{\lower 0.511ex
   \hbox{$\sim$}}}} 
\def\ltap{\mathrel{ \rlap{\raise 0.511ex
   \hbox{$<$}}{\lower 0.511ex \hbox{$\sim$}}}}
   \newcommand{\deltaatm}{\mbox{$\Delta m^2_{31}$}}
   \newcommand{\deltasol}{\mbox{$ \Delta m^2_{21}$}}

   \newcommand{\eV}{\mbox{$ \ \mathrm{eV}$}}

\begin{document}

{\flushright
SISSA 35/04/EP

UCLA/04/TEP/24

hep-ph/0406096

}

\begin{center}
\noindent{\Large \tt \bf 
%------------------------------------------ Title --------------------
Three-Neutrino Oscillations of Atmospheric Neutrinos, 
$\theta_{13}$, Neutrino Mass Hierarchy 
and Iron Magnetized Detectors 
%---------------------------------------------------------------------
}\vspace{4mm}
\renewcommand{\thefootnote}{\fnsymbol{footnote}}

\noindent{\large
%-------------------------------------- Author(s) --------------------
Sergio Palomares-Ruiz$^{1,2}$ and
S.~T.~Petcov$^3$\footnote{Also at: Institute of Nuclear Research and
Nuclear Energy, Bulgarian Academy of Sciences, 1784 Sofia, Bulgaria.}
%---------------------------------------------------------------------
}\vspace{2mm}

\noindent{\small
%------------------------------------ Address(es) --------------------
$^1$ Department of Physics and Astronomy, UCLA,
  Los Angeles, CA 90095, USA
%---------------------------------------------------------------------
\\
%------------------------------------ Address(es) --------------------
$^2$ Department of Physics and Astronomy, Vanderbilt University,
  Nashville, TN 37235, USA
%---------------------------------------------------------------------
\\
%------------------------------------ Address(es) --------------------
$^3$ Scuola Internazionale Superiore di Studi Avanzati 
and Istituto Nazionale di Fisica Nucleare, I-34014 Trieste, Italy}

\end{center}
\vspace{4mm}
\vspace{6mm}
\renewcommand{\thefootnote}{\arabic{footnote}}
\setcounter{footnote}{0}

\begin{abstract}
We derive predictions for the Nadir angle ($\theta_n$)
dependence of the ratio $N_{\mu^-}/N_{\mu^+}$
of the rates of the $\mu-$ and $\mu^+$
multi-GeV events, and for 
the $\mu^{-}-\mu^+$ event 
rate asymmetry,
$A_{\mu^-\mu^+} = [N(\mu^{-}) - N(\mu^+)]/[N(\mu^{-}) + N(\mu^+)]$,
in iron-magnetized calorimeter 
detectors (MINOS, INO, etc.)
in the case of 3-neutrino oscillations
of the atmospheric $\nu_\mu$  and 
$\bar{\nu}_\mu$, driven
by one neutrino mass squared difference,
$|\deltaatm| \sim (2.0 - 3.0)\times 10^{-3}~{\rm eV^2} 
\gg \deltasol$.
The asymmetry $A_{\mu^-\mu^+}$ 
(the ratio $N_{\mu^-}/N_{\mu^+}$) 
is  shown to be particularly sensitive
to the Earth matter effects in the 
atmospheric neutrino oscillations,
and thus to the values
of $\sin^2\theta_{13}$ and
$\sin^2\theta_{23}$, $\theta_{13}$ and
$\theta_{23}$ being the neutrino mixing angle
limited by the CHOOZ and Palo Verde experiments
and that responsible for the dominant 
atmospheric $\nu_\mu \rightarrow \nu_{\tau}$ 
($\bar{\nu}_\mu \rightarrow \bar{\nu}_{\tau}$)
oscillations. It is also very sensitive  
to the type of neutrino mass 
spectrum which can be with normal 
($\deltaatm > 0$) 
or with inverted  
($\deltaatm < 0$) hierarchy.
We find that for  $\sin^2\theta_{23} \gtap 0.50$,
$\sin^22\theta_{13} \gtap 0.06$
and $|\deltaatm| = 
(2 - 3)\times 10^{-3}~{\rm eV^2}$,
the Earth matter effects 
produce a relative difference 
between the {\it integrated asymmetries}
$\bar{A}_{\mu^-\mu^+}$ and $\bar{A}^{2\nu}_{\mu^-\mu^+}$
in the {\it mantle}
($\cos\theta_n = 0.30 - 0.84$)
and {\it core} ($\cos\theta_n = 0.84 - 1.0$) bins,
which is bigger in absolute value than 
approximately $\sim 15\%$,
can reach the values of $(30 - 35)\%$,
and thus can be sufficiently large to be observable.
The sign of the indicated asymmetry difference
is anticorrelated with the sign 
of $\deltaatm$.
An observation of the Earth matter effects 
in the Nadir angle distribution of the 
asymmetry $A_{\mu^-\mu^+}$
(ratio $N_{\mu^-}/N_{\mu^+}$)
would clearly indicate that
$\sin^22\theta_{13} \gtap 0.06$ and 
$\sin^2\theta_{23} \gtap 0.50$,
and would lead to the determination of the
sign of $\deltaatm$.

\end{abstract}
\vspace{20pt}

\newpage
%%%%%%%%%%%%%%%%%%%%%%%%%%%%%%%%%%%%%%%%%%%%%%%%%%%%%%%%%%%%%%%%%%%%%%%%
%			         Introduction			       %
%%%%%%%%%%%%%%%%%%%%%%%%%%%%%%%%%%%%%%%%%%%%%%%%%%%%%%%%%%%%%%%%%%%%%%%%
\section{Introduction}
\hskip 0.6truecm 
 There has been a remarkable  progress in the studies of neutrino
oscillations in the last several years.
The experiments with solar, 
atmospheric and reactor
 neutrinos~\cite{sol,SKsolar,SNO1,SNO2,SNO3,SKatm,KamLAND}  
have provided 
compelling evidences for the 
existence of neutrino oscillations 
driven by nonzero neutrino masses and neutrino mixing.
Evidences for oscillations of neutrinos were
obtained also in the first long baseline
accelerator neutrino experiment K2K~\cite{K2K}.
It was predicted already in 1967~\cite{BPont67}
that the existence of solar neutrino oscillations would 
cause a deficit of solar neutrinos
detected on Earth.
The hypothesis of solar neutrino oscillations, 
which in one variety or another 
were considered starting from the late 60's
as  the most natural explanation 
of the observed ~\cite{sol,SKsolar} 
solar neutrino deficit
(see, e.g., refs.~\cite{BPont67, BiPont78, BiPet87, SPSchlad97}),
has received a convincing
confirmation from the measurement 
of the solar neutrino flux
through the neutral current reaction on
deuterium by the 
SNO experiment~\cite{SNO2,SNO3}, and by the first
results of the KamLAND experiment~\cite{KamLAND}. 
The analysis of the solar neutrino data
obtained by 
Homestake, SAGE, GALLEX/GNO, Super-Kamiokande and SNO
experiments showed that the data favor 
the Large Mixing Angle (LMA) MSW 
solution of the solar neutrino problem 
(see, e.g., ref.~\cite{SNO2}).
The first results of the 
KamLAND reactor experiment~\cite{KamLAND} have confirmed
(under the very plausible assumption of CPT-invariance)
the LMA MSW solution, establishing it essentially as 
a unique solution of the solar neutrino problem. 

   The latest addition to this magnificent effort
is the evidence 
presented recently by the Super-Kamiokande (SK) 
collaboration for an ``oscillation dip'' 
in the $L/E-$dependence,
of the (essentially multi-GeV) 
$\mu-$like atmospheric neutrino events
\footnote{The sample used in the analysis of the
$L/E$ dependence consists of 
$\mu-$like events 
for which the relative uncertainty in the 
experimental determination of the 
$L/E$ ratio does not exceed 70\%.}~\cite{SKdip04}, 
$L$ and $E$ being the 
distance traveled by neutrinos and 
the neutrino energy.
As is well known, the SK atmospheric neutrino data
is best described in terms of dominant 2-neutrino
$\nu_{\mu} \rightarrow \nu_{\tau}$ 
($\bar{\nu}_{\mu} \rightarrow \bar{\nu}_{\tau}$)
vacuum oscillations with maximal mixing,
$\sin^22\theta_{23} \cong 1$.
The observed dip is predicted due to the oscillatory dependence 
of the $\nu_{\mu} \rightarrow \nu_{\tau}$ and
$\bar{\nu}_{\mu} \rightarrow \bar{\nu}_{\tau}$
oscillation probabilities,
$P(\nu_{\mu} \rightarrow \nu_{\tau}) \cong
P(\bar{\nu}_{\mu} \rightarrow \bar{\nu}_{\tau})$,
on $\L/E$.
The dip in the observed $L/E$ distribution
corresponds to the first oscillation minimum
of the $\nu_{\mu}$ ($\bar{\nu}_{\mu}$)
survival probability, 
$P(\nu_{\mu} \rightarrow \nu_{\mu})
= 1 - P(\nu_{\mu} \rightarrow \nu_{\tau})$, 
as $L/E$ increases starting from values for which
$\deltaatm L/(2E) \ll 1$ and 
$P(\nu_{\mu} \rightarrow \nu_{\mu}) \cong 1$,
$\deltaatm$ being the neutrino mass squared difference 
responsible for the atmospheric 
$\nu_{\mu}$ and $\bar{\nu}_{\mu}$ oscillations.
This beautiful result represents the first 
ever observation of a direct effect 
of the oscillatory dependence 
on $L/E$ of the probability 
of neutrino oscillations in vacuum.

    The interpretation of the solar and
atmospheric neutrino, and of  KamLAND
data in terms of 
neutrino oscillations requires
the existence of 3-neutrino mixing
in the weak charged lepton current 
(see, e.g., ref.~\cite{BGG99}):
%%%%%%%%%%%%%%%%%%%
\begin{equation}
\nu_{l \mathrm{L}}  = \sum_{j=1}^{3} U_{l j} \, \nu_{j \mathrm{L}}~.
\label{3numix}
\end{equation}
%%%%%%%%%%%%%%%%%%%
\noindent Here $\nu_{lL}$, $l  = e,\mu,\tau$,
are the three left-handed flavor 
neutrino fields,
$\nu_{j \mathrm{L}}$ is the 
left-handed field of the 
neutrino $\nu_j$ having a mass $m_j$
and $U$ is the Pontecorvo-Maki-Nakagawa-Sakata (PMNS)
neutrino mixing matrix~\cite{BPont57},
%%%%%%%%%%%%%%%%%%%%%%%%%%%%%%%%%%%%%%%%%%%%
\begin{equation}
U = \left(\begin{array}{ccc}
U_{e1}& U_{e2} & U_{e3} \\
U_{\mu 1} & U_{\mu 2} & U_{\mu 3} \\
U_{\tau 1} & U_{\tau 2} & U_{\tau 3} 
\end{array} \right)
= \left(\begin{array}{ccc} 
c_{12}c_{13} & s_{12}c_{13} & s_{13}e^{-i\delta}\\
 - s_{12}c_{23} - c_{12}s_{23}s_{13}e^{i\delta} & 
c_{12}c_{23} - s_{12}s_{23}s_{13}e^{i\delta} & s_{23}c_{13}\\
s_{12}s_{23} - c_{12}c_{23}s_{13}e^{i\delta} 
& -c_{12}s_{23} - s_{12}c_{23}s_{13}e^{i\delta}
& c_{23}c_{13}\\ 
\end{array} \right)
\label{Umix}
\end{equation}

\noindent where we have used a standard parametrization 
of $U$ with the usual notations, $s_{ij} \equiv \sin \theta_{ij}$,
$c_{ij} \equiv \cos \theta_{ij}$, 
and $\delta$ is the Dirac CP-violation phase
\footnote{We have not 
written explicitly the two possible Majorana 
CP-violation phases~\cite{BHP80, Doi81}
which do not enter into the expressions for the oscillation 
probabilities of interest~\cite{BHP80, Lang87}.}.
If one identifies $\Delta m^2_{21} > 0$ and $\Delta m^2_{31}$  
with the neutrino mass squared differences
which drive the solar and atmospheric 
neutrino oscillations, 
the data suggest that 
$|\deltaatm| \gg \deltasol$.
In this case $\theta_{12}$ and $\theta_{23}$, 
represent the neutrino mixing 
angles responsible for the
solar and the dominant atmospheric 
neutrino oscillations,
$\theta_{12}$, 
$\theta_{23}$, 
while $\theta_{13}$ is the angle 
limited by the data from
the CHOOZ and Palo Verde experiments 
\cite{CHOOZ,PaloV}.

  The 3-neutrino oscillations of the solar
$\nu_e$ depend in the case of interest,
$|\deltaatm| \gg \deltasol$,
not only on $\deltasol$ and
$\theta_{12}$,
but also on $\theta_{13}$.
A  combined 3-neutrino oscillation analysis 
of the solar neutrino,
CHOOZ and KamLAND data 
showed~\cite{SNO3BCGPR}
that for $\sin^2 \theta_{13} \ltap 0.05$
the allowed ranges of the solar neutrino oscillation parameters
do not differ substantially 
from those derived in the 
2-neutrino oscillation analyzes
(see, e.g., refs.~\cite{SNO3,SNO3other}).
A description of the 
indicated data in terms of 
$\nu_e \rightarrow \nu_{\mu,\tau}$
and $\bar{\nu}_e \rightarrow \bar{\nu}_{\mu,\tau}$
oscillations is possible (at 99.73\% C.L.)
for $\sin^2\theta_{13}\ltap 0.075$.
The data favor the LMA-I MSW solution 
(see, e.g., refs.~\cite{SNO3BCGPR, SNO3other})
with $\deltasol \cong 
7.2\times 10^{-5}~{\rm eV^2}$ and 
$\sin^2\theta_{12} \cong 0.30$.
The LMA-II solution, corresponding to  
$\deltasol \cong 
1.5\times 10^{-4}~{\rm eV^2}$ and 
approximately the same value of
$\sin^2\theta_{12}$, is severely
constrained by the fact~\cite{MarisSP02} that
the ratio of the rates of the
CC and NC reactions on deuterium,
measured with a relatively high precision
in SNO during the salt phase of 
the experiment~\cite{SNO3},
turned out to be definitely smaller than 0.50.
This solution is currently allowed 
by the data only at 99.13\% C.L.~\cite{SNO3BCGPR}.

  The preliminary results 
of the most recent improved analysis of the
SK atmospheric neutrino data, performed by the
SK collaboration, gave at 90\% C.L.~\cite{SKatmo03}
%%%%%%%%%%%%%%%%%%%%%%%%%%%%%%%%
\begin{equation} 
1.3 \,\times\, 10^{-3}\,\mbox{eV}^2\,\leq\,
|\deltaatm|\, \leq \,3.1\, \times\,10^{-3}\,\mbox{eV}^2~,
~~~~~~~~~~0.90 \leq \sin^22\theta_{23} \leq 1.0,
\label{atmo03a}
\end{equation}
%%%%%%%%%%%%%%%%%%%%%%%%%%%%%%%
\noindent with best fit values $|\deltaatm| = 2.0\times 10^{-3}$
eV$^2$ and $\sin^2 2\theta_{\rm A} = 1.0$.
Adding the K2K data~\cite{K2K}, 
the authors~\cite{Fogliatm0308055}
find the same results for $\sin^22\theta_{\rm A}$, 
the same $|\deltaatm|$
best fit value and
%%%%%%%%%%%%%%%%%%%%%%%%%%%%%%%%
\begin{equation} 
1.55 \,\times\, 10^{-3}\,\mbox{eV}^2\,\ltap\,
|\deltaatm|\, \ltap\,2.60\, \times\,10^{-3}\,\mbox{eV}^2~,
 ~~~~~90\%~{\rm C.L.}
\label{atmo03}
\end{equation}
%%%%%%%%%%%%%%%%%%%%%%%%%%%%%%%
Earlier combined analyzes   
of the SK atmospheric neutrino and the K2K data 
produced somewhat larger values of
$|\deltaatm|$: the  best fit value 
and the 90\% C.L. allowed interval of values,
found, e.g., in ref.~\cite{FogliatmKamL} read
$|\deltaatm| = 2.6 \times 10^{-3} \ \eV$
and   $ 2.0 \times 10^{-3} \, \eV^2 \ltap |\deltaatm| 
\ltap 3.2 \times 10^{-3}\, \eV^2$.
Finally, the values of
$|\deltaatm|$ and  $\sin^22\theta_{23}$,
deduced from the SK analysis of the 
$L/E$ dependence of the observed
$\mu-$like atmospheric neutrino 
events~\cite{SKdip04}, are 
comfortably compatible with the values
obtained in the other analyzes:
$|\deltaatm| = 2.4 \times 10^{-3} \ \eV$,
$\sin^22\theta_{23} = 1$ (best fit), and
$1.9 \times 10^{-3} \, \eV^2 \ltap |\deltaatm| 
\ltap 3.0 \times 10^{-3}\, \eV^2$,
$0.90 \leq \sin^22\theta_{23} \leq 1.0$
(90\% C.L.). We will use in our further
discussion  as illustrative the values 
$|\deltaatm| = (2.0;~3.0)\times 10^{-3}~{\rm eV^2}$
and $\sin^22\theta_{23} = 0.92;~ 1.0$.

   Let us note that the 
atmospheric neutrino and K2K data
do not allow one to determine the signs
of $\deltaatm$, and of 
$\cos2\theta_{23}$ if
$\sin^22\theta_{23} \neq 1.0$.
This implies that 
in the case of 3-neutrino mixing 
one can have $\deltaatm > 0$
or $\deltaatm < 0$. The two 
possibilities correspond to two different
types of neutrino mass spectrum:
with normal hierarchy (NH),
$m_1 < m_2 < m_3$, and 
with inverted hierarchy (IH),
$m_3 < m_1 < m_2$.  
The fact that the sign of 
$\cos2\theta_{23}$ 
is not determined when
$\sin^22\theta_{23} \neq 1.0$ implies
that when, e.g., 
$\sin^22\theta_{23} = 0.92$,
two values of  $\sin^2\theta_{23}$ are possible,
$\sin^2\theta_{23} \cong 0.64~{\rm or}~ 0.36$.

  The precise limit on the angle $\theta_{13}$
from the CHOOZ and Palo Verde
data is $\deltaatm-$ dependent 
(see, e.g, ref.~\cite{BNPChooz}).
Using the 99.73\%  allowed range of 
$\deltaatm = (1.1 - 3.2)\times 10^{-3}~{\rm eV^2}$,
from ref.~\cite{Fogliatm0308055}, 
one gets 
from  a combined 3-neutrino
oscillation analysis of the solar neutrino,
CHOOZ and KamLAND data~\cite{SNO3BCGPR}:
%%%%%%%%%%%%%%%%%%%%%%%
\begin{equation}
\sin^2 \theta_{13} < 0.047~(0.074),~~~~~~~~~ 90\%~(99.73\%)~{\rm C.L.}
\end{equation}
%%%%%%%%%%%%%%%%%%%%%%%%%%%%%%%%%%%%
The global analysis of the
solar, atmospheric and reactor neutrino data 
performed in ref.~\cite{ConchaNOON04}
gives $\sin^2 \theta_{13} < 0.054$ at 99.73\% C.L.

 It is difficult to overestimate 
the importance of 
getting more precise 
information about the value 
of the mixing angle $\theta_{13}$,
of determining the sign of $\deltaatm$, or
the type of the neutrino mass spectrum 
(with normal or inverted hierarchy), 
and of measuring the value of 
$\sin^2\theta_{23}$ with a 
higher precision,
for the future progress in the 
studies of neutrino mixing.
Although this has been widely recognized, let us 
repeat the arguments on which the statement is based.

   The mixing angle $\theta_{13}$, or
the absolute value
of the element $U_{e3}$ of the PMNS matrix,
$|U_{e3}| =\sin\theta_{13}$, 
plays a very important  
role in the phenomenology of 
the 3-neutrino oscillations.
It drives the sub-dominant 
 $\nu_{\mu} \leftrightarrow \nu_e$
($\bar{\nu}_{\mu} \leftrightarrow \bar{\nu}_e$)
oscillations of the atmospheric 
$\nu_{\mu}$ ($\bar{\nu}_{\mu}$) and
$\nu_e$ ($\bar{\nu}_e$)~\cite{SP3198,SPNu98}. 
The value of $\theta_{13}$   
controls also the 
$\nu_{\mu} \rightarrow \nu_e$,
$\bar{\nu}_{\mu} \rightarrow \bar{\nu}_e$,
$\nu_{e} \rightarrow \nu_{\mu}$ and
$\bar{\nu}_{e} \rightarrow \bar{\nu}_{\mu}$
transitions in the long baseline neutrino 
oscillation experiments (MINOS, CNGS),
and in the widely discussed
very long baseline neutrino oscillation 
experiments at neutrino factories (see, e.g.,
refs.~\cite{LBL,AMMS99,mantle}).  
The magnitude of the 
T-violating and CP-violating terms 
in neutrino oscillations probabilities 
is directly proportional
to $\sin\theta_{13}$ (see, e.g., refs.~\cite{3nuKP88,CPother}).

 If the neutrinos with definite mass 
are Majorana particles (see, e.g., ref.~\cite{BiPet87}),
the predicted value of the effective Majorana mass
parameter in neutrinoless double $\beta-$decay
depends strongly in the case of normal hierarchical
or partially hierarchical
neutrino mass spectrum 
on the value of $\sin^2\theta_{13}$ 
(see, e.g., ref.~\cite{BPP1}).   

   The sign of $\deltaatm$ determines, for instance,
which of the transitions (e.g., of atmospheric neutrinos)
$\nu_{\mu} \rightarrow \nu_e$ and
$\nu_{e} \rightarrow \nu_{\mu}$, or
$\bar{\nu}_{\mu} \rightarrow \bar{\nu}_e$ and
$\bar{\nu}_{e} \rightarrow \bar{\nu}_{\mu}$,
can be enhanced by the Earth matter 
effects~\cite{LW78,BPPW80,MS85}.
The predictions for the
neutrino effective Majorana mass
in neutrinoless double $\beta-$decay
depend critically on the type of the
neutrino mass spectrum 
(normal or inverted hierarchical)~\cite{BPP1,PPSNO23bb}.
The knowledge of the value of $\theta_{13}$ 
and of the sign of $\deltaatm$
is crucial for the searches 
for the correct theory of 
neutrino masses and mixing as well.

  Somewhat better limits on $\sin^2 \theta_{13}$ than 
the existing one can be obtained in the 
MINOS, OPERA and ICARUS experiments~\cite{MINOS,OPICA03}. 
Various options are being currently discussed
(experiments with off-axis neutrino beams, more precise
reactor antineutrino and long baseline experiments, etc.,
see, e.g., ref.~\cite{MSpironu02}) of how to improve
by at least an order of magnitude, i.e., 
to values of $\sim 0.005$ or smaller, 
the sensitivity to $\sin^2\theta_{13}$. 
The sign of $\deltaatm$ can be determined in
very long baseline neutrino oscillation 
experiments at neutrino factories
(see, e.g., refs.~\cite{LBL,AMMS99}), and, e.g, using
combined data from long baseline
oscillation experiments at the JHF facility and
with off-axis neutrino beams~\cite{HLM}.
If the neutrinos with definite mass are
Majorana particles, it can be determined
by measuring the effective neutrino Majorana
mass in neutrinoless double $\beta-$decay 
experiments~\cite{BPP1,PPSNO23bb}.
Under certain rather special
conditions it might be determined also
in experiments with reactor 
$\bar{\nu}_e$~\cite{SPMPiai01}. 

   In the present article we study possibilities
to obtain information on the value 
of $\sin^2\theta_{13}$ and on the sign of
$\deltaatm$
using the data on atmospheric 
neutrinos, which can be obtained in experiments
with detectors able to measure the charge of the muon
produced in the charged current (CC) reaction
by atmospheric $\nu_{\mu}$ or
$\bar{\nu}_{\mu}$. It is a natural continuation of
our similar study
for water-\v{C}erenkov detectors~\cite{JBSP203}. 
In the experiments with muon 
charge identification it will 
be possible to distinguish between 
the $\nu_{\mu}$ and $\bar{\nu}_{\mu}$ induced events.
As is well known, the water-\v{C}erenkov detectors
do not have such a capability.
Among the operating detectors, MINOS
has muon charge identification
capabilities for multi-GeV 
muons~\cite{MINOS}.
The MINOS experiment is currently collecting 
atmospheric neutrino data. 
The detector has relatively small mass,
but after 5 years of data-taking 
it is expected to collect about 440 
atmospheric $\nu_{\mu}$ 
and about 260 atmospheric $\bar{\nu}_{\mu}$ 
multi-GeV events (having the interaction 
vertex inside the detector).
There are also plans to build a 30-50 kton
magnetized tracking iron calorimeter 
detector in India
within the India-based 
Neutrino Observatory (INO) 
project~\cite{INO}.
The INO detector will be based on  
MONOLITH design~\cite{MONOLITH}. 
The primary goal is to study
the oscillations of the 
atmospheric $\nu_{\mu}$ and $\bar{\nu}_{\mu}$.
This detector is planned to have
efficient muon charge identification,
high muon energy resolution ($\sim 5\% $) and 
muon energy threshold of about 2 GeV.
It will accumulate sufficiently high statistics
of atmospheric $\nu_{\mu}$ and $\bar{\nu}_{\mu}$
induced events in several years, which would permit to 
search for effects of the subdominant
$\nu_{\mu} \rightarrow \nu_e$ 
($\nu_{e} \rightarrow \nu_{\mu}$) and
$\bar{\nu}_{\mu} \rightarrow \bar{\nu}_e$
($\bar{\nu}_{e} \rightarrow \bar{\nu}_{\mu}$)
transitions.

 If $\nu_{\mu}$ and $\bar{\nu}_{\mu}$
with energies $E_{\nu,\bar{\nu}} \gtap$ 2 GeV
take part in 2-neutrino
$\nu_{\mu} \rightarrow \nu_{\tau}$ and
$\bar{\nu}_{\mu} \rightarrow \bar{\nu}_{\tau}$
oscillations, one would have 
$P(\nu_{\mu} \rightarrow \nu_{\tau}) =  
P(\bar{\nu}_{\mu} \rightarrow \bar{\nu}_{\tau})$.
If a given detector collects
a sample of atmospheric $\nu_{\mu}$ and $\bar{\nu}_{\mu}$
induced CC events with $\mu^{\pm}$ having energies 
exceeding a few GeV, 
the difference between the
$\mu^{-}$ and $\mu^{+}$ event rates 
in magnetized iron detectors
of interest (MINOS, INO), apart from 
detector effects, would essentially be 
due to the difference
between the $\nu_{\mu}$ and $\bar{\nu}_{\mu}$
charged current (CC) deep inelastic
scattering cross sections.
In the case of
3-neutrino oscillations 
of the atmospheric $\nu_{\mu}$, $\bar{\nu}_{\mu}$,
$\nu_e$ and $\bar{\nu}_e$, the Earth matter 
effects can generate a difference between the  
$\nu_{\mu} \rightarrow \nu_e$ 
($\nu_{e} \rightarrow \nu_{\mu}$) and
$\bar{\nu}_{\mu} \rightarrow \bar{\nu}_e$
($\bar{\nu}_{e} \rightarrow \bar{\nu}_{\mu}$)
transitions. For 
$\deltasol \ll |\deltaatm|$,
which is implied 
by the current solar and atmospheric 
neutrino data,
and if $\sin^2\theta_{13}\neq 0$, 
the Earth matter effects 
can resonantly enhance either
the $\nu_{\mu} \rightarrow \nu_e$ and
$\nu_{e} \rightarrow \nu_{\mu}$,
or the $\bar{\nu}_{\mu} \rightarrow \bar{\nu}_e$
and $\bar{\nu}_{e} \rightarrow \bar{\nu}_{\mu}$
transitions, depending on the sign of 
$\deltaatm$.
The effects of the enhancement 
can be substantial for $\sin^2\theta_{13}\gtap 0.01$.
Under the condition
$\deltasol \ll |\deltaatm|$,
the expressions for the 
probabilities of 
$\nu_{\mu} \rightarrow \nu_e$ 
($\nu_{e} \rightarrow \nu_{\mu}$) and
$\bar{\nu}_{\mu} \rightarrow \bar{\nu}_e$
($\bar{\nu}_{e} \rightarrow \bar{\nu}_{\mu}$)
transitions contain also 
$\sin^2\theta_{23}$ as a factor
which determines their maximal values. 
Since the fluxes of multi-GeV
$\nu_{\mu}$ and  $\nu_e$ 
($\bar{\nu}_{\mu}$ and $\bar{\nu}_e$)
differ considerably (by a factor $\sim (2.6 - 4.5)$ for
$E_{\nu,\bar{\nu}} \sim (2 - 10)$ GeV),
the Earth matter effects 
can create a substantial difference between
the $\mu^-$  and $\mu^+$ rates of events,
produced by the atmospheric 
$\nu_{\mu}$ and $\bar{\nu}_{\mu}$
in MINOS, INO, or any other detector of the same type.
This difference can be relatively large 
and  observable 
in the samples of the 
multi-GeV $\mu^{\pm}$ events ($E_{\mu} \sim (2 - 10)$ GeV), 
in which the muons
are produced by atmospheric 
neutrinos with relatively large path length
in the Earth, i.e., by neutrinos
crossing deeply the 
Earth mantle~\cite{AMMS99,mantle}
or the mantle and the core~\cite{SP3198,107,core,ConchaMal02}.
These are the neutrinos for which~\cite{JBSP203}
$\cos\theta_n \gtap (0.3 - 0.4)$,
$\theta_n$ being the Nadir angle
characterizing the (atmospheric) 
neutrino trajectory in the Earth.

  For $\deltaatm > 0$,    
the $\nu_{\mu} \rightarrow \nu_e$ 
($\bar{\nu}_{\mu} \rightarrow \bar{\nu}_e$)
and $\nu_{e} \rightarrow \nu_{\mu}$
($\bar{\nu}_{e} \rightarrow \bar{\nu}_{\mu}$)
transitions in the Earth lead to a reduction
of the rate of the multi-GeV $\mu^{-}$
events 
observable in MINOS, INO, etc.,
with respect to the case of absence
of these transitions (see, e.g., 
refs.~\cite{SP3198,106,107,core,Taba02}). If 
$\deltaatm < 0$,
the  
$\mu^{+}$ event rate will be reduced.
Correspondingly, as observable which is sensitive 
to the Earth matter effects, and thus to the value
of $\sin^2\theta_{13}$ and the sign of 
$\deltaatm$, as well as to 
$\sin^2\theta_{23}$, we can consider the
Nadir-angle distribution of the 
ratio $N(\mu^{-})/N(\mu^+)$ 
of the multi-GeV $\mu^-$ and $\mu^+$ 
event rates, or, equivalently, of the $\mu^- - \mu^+$ 
event rate asymmetry
$A_{\mu^-\mu^+} = [N(\mu^{-}) - N(\mu^+)]/[N(\mu^{-}) + N(\mu^+)]$.
The systematic uncertainty, in particular,
in the Nadir angle dependence of 
$N(\mu^{-})/N(\mu^+)$, and correspondingly in 
the asymmetry $A_{\mu^-\mu^+}$, can 
be smaller than those on the measured
Nadir angle distributions of the rates of
$\mu^-$ and $\mu^+$ events,
$N(\mu^-)$ and $N(\mu^+)$.

  We  have obtained predictions
for the Nadir-angle distribution of 
$A_{\mu^-\mu^+}$
in the case of 3-neutrino
oscillations of the atmospheric 
$\nu_{\mu}$, $\bar{\nu}_{\mu}$,
$\nu_e$ and $\bar{\nu}_e$,
both for neutrino mass spectra with 
normal ($\deltaatm > 0$)
and inverted  ($\deltaatm < 0$) hierarchy,
$(A^{3\nu}_{\mu^-\mu^+})_{\rm NH}$, 
$(A^{3\nu}_{\mu^-\mu^+})_{\rm IH}$,
and for $\sin^2\theta_{23} = 0.64;~0.50;~0.36$.
These are compared  with the predicted
Nadir-angle distributions of the
same asymmetry in the case of 2-neutrino
$\nu_{\mu} \rightarrow \nu_{\tau}$ and
$\bar{\nu}_{\mu} \rightarrow \bar{\nu}_{\tau}$
oscillations  of the atmospheric $\nu_{\mu}$ and 
$\bar{\nu}_{\mu}$ (i.e., for $\sin^2\theta_{13} = 0$),
$A^{2\nu}_{\mu^-\mu^+}$.
Predictions for the three types of 
asymmetries indicated 
above of the suitably integrated
Nadir angle distributions
of the $\mu^-$ and $\mu^+$ 
multi-GeV event rates are also given. 
Our results show, in particular,
that for $\sin^2\theta_{23} \gtap 0.50$
and $\sin^22\theta_{13} \gtap 0.06$ 
the effects of the Earth matter enhanced
subdominant transitions 
of the atmospheric neutrinos,
$\nu_{\mu} \rightarrow \nu_e$ and 
and $\nu_{e} \rightarrow \nu_{\mu}$,
or $\bar{\nu}_{\mu} \rightarrow \bar{\nu}_e$ 
and $\bar{\nu}_{e} \rightarrow \bar{\nu}_{\mu}$,
can be sufficiently large to be 
observable with INO and possibly with MINOS detectors.
Conversely, if the indicated effects 
are observed in the MINOS and/or INO experiments,
that would imply that $\sin^22\theta_{13} \gtap 0.05$, 
$\sin^2\theta_{23} \gtap 0.50$ and 
at the same time would permit
to determine the sign of
$\deltaatm$ and thus to answer  
the fundamental question about the type of
hierarchy - normal or inverted, the neutrino mass
spectrum has. 

 Let us note that the Earth matter effects 
in atmospheric neutrino oscillations
have been widely studied (for recent detailed 
analyzes see, e.g., refs.~\cite{ConchaMal02,JBSP203},  
which contain also a rather complete list 
of references to earlier work on the subject). 
A rather detailed analysis for
the MONOLITH detector has been performed in 
ref.~\cite{Taba02}. 
A large number of studies 
have been done for the Super-Kamiokande 
detector, or more generally, 
for water-\v{C}erenkov detectors.
In ref.~\cite{JBSP203}, in particular, the magnitude of the
Earth matter effects in the Nadir angle
distribution of the ratio of the multi-GeV
$\mu-$like and $e-$like events, 
measured in  water-\v{C}erenkov 
detectors, $N_{\mu}/N_e$,
has been investigated.
This Nadir angle distribution 
is the observable 
most sensitive to the matter 
effects of interest.
It was concluded that for $\sin^2\theta_{23} \gtap 0.50$,
$\sin^2\theta_{13} \gtap 0.01$ and
$\deltaatm > 0$,
the effects of the Earth matter enhanced
$\nu_{\mu} \rightarrow \nu_e$ 
and $\nu_{e} \rightarrow \nu_{\mu}$
transitions of the atmospheric 
$\nu_{\mu}$ and $\nu_e$,
might be observable with the 
Super-Kamiokande detector.
However, determining the sign of 
$\deltaatm$ would be 
quite challenging in this experiment
(or its bigger version - 
Hyper-Kamiokande~\cite{HyperK}). 
In general, the matter effects 
in the Nadir angle distribution
of the ratio of the multi-GeV $\mu-$like and $e-$like
events, $N_{\mu}/N_e$, which can be measured in the
Super-Kamiokande or  other water-\v{C}erenkov detectors,
are smaller than the matter effects 
in the Nadir angle distribution
of the ratio of the multi-GeV $\mu^-$ and $\mu^+$
events, $N(\mu^-)/N(\mu^+)$,
which can be measured in MINOS, INO, or 
any other atmospheric neutrino experiment with
sufficiently good muon charge identification. 
The reason is that in the case of 
water-\v{C}erenkov detectors, 
approximately 2/3 of the rate of the multi-GeV
$\mu-like$ events is due to $\nu_{\mu}$, 
and $\sim 1/3$ is due to $\bar{\nu}_{\mu}$;
similar partition is valid for the
multi-GeV $e-$like events.
Depending on the sign of $\deltaatm$,
the matter effects enhance either the
neutrino transitions,
$\nu_{\mu} \rightarrow \nu_e$ and 
and $\nu_{e} \rightarrow \nu_{\mu}$,
or the antineutrino transitions,
 $\bar{\nu}_{\mu} \rightarrow \bar{\nu}_e$ 
and $\bar{\nu}_{e} \rightarrow \bar{\nu}_{\mu}$,
but not both types of transitions.
Correspondingly, because the
$\nu_{\mu,e}-$ and $\bar{\nu}_{\mu,e}-$ 
induced events are indistinguishable
in water-\v{C}erenkov detectors,
only $\sim 2/3$ or $\sim 1/3$ 
of the events in the
multi-GeV $\mu-$like and $e-$like
samples collected in these detectors
are due to neutrinos whose transitions
can be enhanced by matter effects.
This effectively reduces the magnitude 
of the matter effects in the 
samples of multi-GeV $\mu-$like and $e-$like events.
Obviously, such a ``dilution''
of the magnitude of the matter effects 
does not take place in the  
samples of the multi-GeV $\mu^-$ and $\mu^+$ events,
which can be collected
in MINOS and INO experiments, i.e.,
in the experiments with muon charge
identification.

%%%%%%%%%%%%%%%%%%%%%%%%%%%%%%%%%%%%%%%%%%%%%%%%%%%%%%%%%%%%%%%%%%%%%%%%
\section{Subdominant 3-$\nu$ Oscillations of Multi-GeV
Atmospheric Neutrinos in the Earth}
%%%%%%%%%%%%%%%%%%%%%%%%%%%%%%%%%%%%%%%%%%%%%%%%%%%%%%%%%%%%%%%%%%%%%%%%

\hskip 0.6truecm 
In the present Section we review
the physics of the subdominant 3-neutrino 
oscillations of the multi-GeV 
atmospheric neutrinos in the Earth 
(see, e.g., ref.~\cite{JBSP203}).

   The subdominant $\nu_{\mu} \rightarrow \nu_{e}$
($\bar{\nu}_{\mu} \rightarrow \bar{\nu}_{e}$)
and $\nu_{e} \rightarrow \nu_{\mu (\tau)}$
($\bar{\nu}_{e} \rightarrow \bar{\nu}_{\mu (\tau)}$)
oscillations of the multi-GeV 
atmospheric neutrinos of interest should exist 
and their effects could be observable 
if three-flavor-neutrino 
mixing takes place in vacuum,
i.e., if $\sin^22\theta_{13} \neq 0$, and if $\sin^22\theta_{13}$ is
sufficiently large~\cite{SP3198,core,ConchaMal02}.  
These transitions are 
driven by $\deltaatm$.
The probabilities of the transitions contain
$\sin^2\theta_{23}$ as factor which determines
their maximal value (see further).
For $\deltaatm > 0$, the
 $\nu_{\mu} \rightarrow \nu_{e}$ 
($\bar{\nu}_{\mu} \rightarrow \bar{\nu}_{e}$)
and $\nu_{e} \rightarrow \nu_{\mu (\tau)}$
($\bar{\nu}_{e} \rightarrow \bar{\nu}_{\mu (\tau)}$)
transitions of the multi-GeV 
atmospheric neutrinos, are amplified (suppressed) 
by the Earth matter effects; if
$\deltaatm < 0$, the transitions of 
neutrinos are suppressed and those of antineutrinos
are enhanced. Therefore for a given sign of 
$\deltaatm$, the Earth matter affects 
differently the transitions of neutrinos and antineutrinos.
Thus, the study of the subdominant atmospheric neutrino
oscillations can provide information, 
in particular, about the type of 
neutrino mass hierarchy (or sign of $\deltaatm$), the magnitude of 
$\sin^2\theta_{13}$ and the octant where $\theta_{23}$ lies.

  Under the condition
$|\deltaatm| \gg \deltasol$, which 
the neutrino mass squared differences 
determined from the existing atmospheric and 
solar neutrino and KamLAND data satisfy, 
the relevant 3-neutrino 
$\nu_{\mu} \rightarrow \nu_{e}$ 
($\bar{\nu}_{\mu} \rightarrow \bar{\nu}_{e}$)
and $\nu_{e} \rightarrow \nu_{\mu (\tau)}$
($\bar{\nu}_{e} \rightarrow \bar{\nu}_{\mu (\tau)}$)
transition probabilities reduce effectively 
to a 2-neutrino transition probability~\cite{3nuSP88} with 
$\deltaatm$
and $\theta_{13}$ playing the role of the relevant
2-neutrino oscillation parameters.

    The 3-neutrino oscillation probabilities 
of interest for atmospheric $\nu_{e,\mu}$ having energy $E$
and crossing the Earth along a trajectory characterized by a 
Nadir angle $\theta_{n}$,
have the following form~\cite{3nuSP88}:
%%%%%%%%%%%%%%%%%
\begin{equation}
P_{3\nu}(\nu_{e} \rightarrow \nu_{e}) \cong 1 -
P_{2\nu}, 
\label{P3ee}
\end{equation}
\begin{equation}
P_{3\nu}(\nu_{e} \rightarrow \nu_{\mu}) \cong
P_{3\nu}(\nu_{\mu} \rightarrow \nu_{e}) \cong
s_{23}^2~P_{2\nu},
\label{P3emu}
\end{equation}
\begin{equation}
P_{3\nu}(\nu_{e} \rightarrow \nu_{\tau}) \cong
c_{23}^2~P_{2\nu},
\label{P3etau}
\end{equation}
\begin{equation}
P_{3\nu}(\nu_{\mu} \rightarrow \nu_{\mu}) \cong 1 -
s_{23}^4~P_{2\nu}
- 2c^2_{23}s^2_{23}~\left [ 1 -
Re~( e^{-i\kappa}
A_{2\nu}(\nu_{\tau} \rightarrow \nu_{\tau}))\right ] ,
\label{P3mumu}
\end{equation}
\begin{equation}
P_{3\nu}(\nu_{\mu} \rightarrow \nu_{\tau }) = 
1 - P_{3\nu}(\nu_{\mu} \rightarrow \nu_{\mu}) - 
P_{3\nu}(\nu_{\mu} \rightarrow \nu_{e}).
\label{P3mutau}
\end{equation}
%%%%%%%%%%%%%%%%%%%%%%%%%%%
\noindent  Here $P_{2\nu} \equiv 
P_{2\nu}(\deltaatm, \theta_{13};E,\theta_{n})$
is the probability of 2-neutrino 
$\nu_{e} \rightarrow \nu'_{\tau}$
oscillations in the Earth,
where  $\nu'_{\tau} = s_{23}\nu_{\mu} + c_{23}
\nu_{\tau}$~\cite{3nuSP88}, 
and $\kappa$ and 
$A_{2\nu}(\nu_{\tau} \rightarrow \nu_{\tau}) \equiv A_{2\nu}$
are known phase and 
2-neutrino transition probability
amplitude~\cite{SP3198,SPNu98,JBSP203}. 
 
The fluxes of atmospheric $\nu_{e,\mu}$ 
of energy $E$, which reach the detector after
crossing the Earth along a given trajectory  
specified by the value of $\theta_{n}$, 
$\Phi_{\nu_{e,\mu}}(E,\theta_{n})$, 
are given by the following expressions 
in the case of the 3-neutrino oscillations  
under discussion~\cite{SPNu98}:
%%%%%%%%%%%%%%%%%%%%%%%%%%%%%%%%%%%%%%%
\begin{equation}
\Phi_{\nu_e}(E,\theta_{n}) \cong 
\Phi^{0}_{\nu_e}~\left (  1 + 
  [s^2_{23}r - 1]~P_{2\nu}\right ),
\label{Phie}
\end{equation}
\begin{equation}
\Phi_{\nu_{\mu}}(E,\theta_{n}) \cong \Phi^{0}_{\nu_{\mu}} 
\left ( 1 +
 s^4_{23}~ [(s^2_{23}~r)^{-1} - 1]~P_{2\nu}
 - 2c^2_{23}s^2_{23}~\left [ 1 -
Re~( e^{-i\kappa}
A_{2\nu}(\nu_{\tau} \rightarrow \nu_{\tau})) \right ] \right )~, 
\label{Phimu}
\end{equation}
%%%%%%%%%%%%%%%%%%%%%%%
\noindent where $\Phi^{0}_{\nu_{e(\mu)}} = 
\Phi^{0}_{\nu_{e(\mu)}}(E,\theta_{n})$ is the
$\nu_{e(\mu)}$ flux in the absence of neutrino 
oscillations and
\vspace{-0.4cm}
%%%%%%%%%%%%%%%%%
\begin{equation}
r \equiv r(E,\theta_{n}) \equiv
\frac{\Phi^{0}_{\nu_{\mu}}(E,\theta_{z})} 
{\Phi^{0}_{\nu_{e}}(E,\theta_{z})}~~.
\label{r}
\end{equation}
%%%%%%%%%%%%%%%%%%%
\indent  The interpretation of the 
SK atmospheric 
neutrino data in terms of $\nu_{\mu} \rightarrow \nu_{\tau}$
oscillations requires the parameter 
$s^2_{23}$ to lie approximately in the interval
(0.30 - 0.70), with 0.5 being the statistically 
preferred value. For the predicted
ratio $r(E,\theta_{n})$ of the atmospheric 
$\nu_{\mu}$ and $\nu_e$ fluxes 
for i) the Earth core crossing and ii) only 
mantle crossing neutrinos, 
having trajectories for which
$0.3 \ltap \cos\theta_{n}\leq 1.0$, one has 
\cite{Honda,Bartol,Naumov}:
$r(E,\theta_{z}) \cong (2.0 - 2.5)$ for 
the neutrinos giving  
contribution to the sub-GeV
samples of Super-Kamiokande events, 
and $r(E,\theta_{n}) \cong (2.6 - 4.5)$ for those 
giving the main contribution to the multi-GeV samples.
If $s^2_{23} = 0.5$ and $r(E,\theta_{z}) \cong 2.0$,
we have $(s^2_{23}~r(E,\theta_{z}) - 1) \cong 0$,
$((s^2_{23}~r(E,\theta_{z}))^{-1} - 1) \cong 0$,
and the possible effects of the 
$\nu_{\mu} \rightarrow \nu_{e}$ 
and $\nu_{e} \rightarrow \nu_{\mu (\tau)}$ 
transitions on the $\nu_e$ and $\nu_{\mu}$ 
fluxes, and correspondingly in 
the sub-GeV $e-$like and $\mu-$like 
samples of events, 
would be strongly suppressed
independently of the values of the 
corresponding transition probabilities.
For the multi-GeV neutrinos one finds
$s^4_{23}[1 - (s^2_{23}~r(E,\theta_{z}))^{-1}] \cong 
0.06 - 0.14~(0.16 - 0.27)$
and $(s^2_{23}~r(E,\theta_{z}) - 1) 
\cong 0.3 - 1.3~(0.66 - 1.9)$ for 
$s^2_{23} = 0.5~(0.64)$.  
Obviously, the effects of interest 
are much larger for the
multi-GeV neutrinos 
than for the sub-GeV neutrinos. They are also 
predicted to be larger for
the flux of (and event rate due to)
multi-GeV  atmospheric $\nu_e$
than for the flux of (and event rate due to) 
multi-GeV atmospheric $\nu_\mu$.

  The same conclusions are valid for the 
effects of oscillations on the fluxes of, and 
event rates due to, atmospheric antineutrinos
$\bar{\nu}_e$ and $\bar{\nu}_{\mu}$.
The formulae for anti-neutrino 
fluxes and oscillation probabilities
are analogous to those for neutrinos: they can be obtained
formally from eqs.~(\ref{P3ee}) -~(\ref{r})
by replacing the neutrino related quantities - 
probabilities, $\kappa$,
$A_{2\nu}(\nu_{\tau} \rightarrow \nu_{\tau})$ 
and fluxes, with the 
corresponding quantities for antineutrinos:
$P_{2\nu}(\deltaatm, \theta_{13};E,\theta_{n})
\rightarrow \bar{P}_{2\nu}(\deltaatm, \theta_{13};E,\theta_{n})$,
$\kappa \rightarrow \bar{\kappa}$, 
$A_{2\nu}(\nu_{\tau} \rightarrow \nu_{\tau}) \rightarrow
A_{2\nu}(\bar{\nu}_{\tau} \rightarrow \bar{\nu}_{\tau}) 
\equiv \bar{A}_{2\nu}$,
$P_{3\nu}(\nu_{l} \rightarrow \nu_{l'}) \rightarrow
P_{3\nu}(\bar{\nu}_{l} \rightarrow \bar{\nu}_{l'})$,
$\Phi^{(0)}_{\nu_{e,\mu}}(E,\theta_{n}) \rightarrow 
\Phi^{(0)}_{\bar{\nu}_{e,\mu}}(E,\theta_{n})$ and
$r(E,\theta_{n}) \rightarrow \bar{r}(E,\theta_{n})$
(see refs.~\cite{SPNu98,JBSP203}).

  Equations~(\ref{P3ee}) -~(\ref{P3mumu}),~(\ref{Phie}) -~(\ref{Phimu}) 
and the similar equations for
antineutrinos imply that in the case under study 
the effects of the $\nu_{\mu} \rightarrow \nu_{e}$,
$\bar{\nu}_{\mu} \rightarrow \bar{\nu}_{e}$, 
and $\nu_{e} \rightarrow \nu_{\mu (\tau)}$,
$\bar{\nu}_{e} \rightarrow \bar{\nu}_{\mu (\tau)}$,
oscillations 
i) increase with the increase of $s^2_{23}$ and are maximal
for the largest allowed value of $s^2_{23}$,
ii) should be substantially larger in the multi-GeV 
samples of events than in the sub-GeV samples, and
iii) in the case of the multi-GeV samples, for 
$\deltaatm > 0$
they lead to a decrease of the 
$\mu^{-}$ event rate,
while if 
$\deltaatm < 0$,  
the $\mu^{+}$ event rate
will  decrease.
The last point follows from the fact that
the magnitude of the effects we are interested in 
depends also on the 2-neutrino oscillation probabilities,
$P_{2\nu}$ and $\bar{P}_{2\nu}$,
and that $P_{2\nu}$ or
$\bar{P}_{2\nu}$ (but not both probabilities) 
can be strongly enhanced  by the Earth matter effects. 
In the case of oscillations in vacuum we have
$P_{2\nu} = \bar{P}_{2\nu} \sim \sin^22\theta_{13}$. 
Given the existing limits on
$\sin^22\theta_{13}$, the 
probabilities $P_{2\nu}$ and
$\bar{P}_{2\nu}$ cannot be large if the oscillations 
take place in vacuum. 

 If $\sin^2\theta_{13}\neq 0$, 
the Earth matter effects 
can resonantly enhance either
the $\nu_{\mu} \rightarrow \nu_e$ and
$\nu_{e} \rightarrow \nu_{\mu}$,
or the $\bar{\nu}_{\mu} \rightarrow \bar{\nu}_e$
and $\bar{\nu}_{e} \rightarrow \bar{\nu}_{\mu}$
transitions, depending on the sign of 
$\deltaatm$. The enhancement mechanisms 
are discussed briefly in the next subsection. 

%%%%%%%%%%%%%%%%%%%%%%%%%%%%%%%%%%%%
\subsection{Enhancing Mechanisms}
%%%%%%%%%%%%%%%%%%%%%%%%%%%%%%%%%%%%%

\hskip 0.6truecm  
As is well-known, the Earth 
density distribution
in the existing Earth models is assumed to be 
spherically symmetric
\footnote{Let us note that 
because of the approximate 
spherical symmetry of the Earth, 
a given neutrino trajectory through 
the Earth is completely 
specified by its Nadir angle.}
and there are two major density structures - 
the core and the mantle, and 
a certain number of substructures (shells or layers).
The core radius and the depth of the mantle
are known with a rather good precision 
 and these data are incorporated
in the Earth models.
According to the Stacey 1977 and the more recent PREM
models~\cite{Stacey:1977,PREM81}, which are widely used in the  
calculations of the probabilities of neutrino oscillations
in the Earth, the core has a radius $R_c = 3485.7~$km,
the Earth mantle depth is approximately $R_{man} = 2885.3~$km,
and the Earth radius is $R_{\oplus} = 6371~$km.  
The mean values of the matter densities and the electron fraction 
numbers in the mantle and in the core read, respectively: 
$\bar{\rho}_{man} \cong 4.5~{\rm g/cm^3}$, 
$\bar{\rho}_c \cong 11.5~{\rm g/cm^3}$, and~\cite{Art2}
$Y_e^{man} = 0.49$,  $Y_e^{c} = 0.467$. The corresponding
mean electron number densities 
in the mantle and in the core read:
$\bar{N}_{e}^{man} = \bar{\rho}_{man} Y_e^{man}/m_{N} \cong
2.2~N_Acm^{-3}$, 
$\bar{N}_{e}^{c}  = \bar{\rho}_{c} Y_e^{c}/m_{N} \cong 5.4~N_Acm^{-3}$,
$m_N$ and $N_A$ being the nucleon mass and 
Avogadro number. 

   Numerical calculations show~\cite{SP3198,3nuKP88} 
that, e.g., the 
$\nu_{e} \rightarrow \nu_{\mu}$
oscillation probability of interest,
calculated within 
the two-layer model of the Earth with 
$\bar{\rho}_{man}$ (or $\bar{N}_{e}^{man}$)
and $\bar{\rho}_{c}$ (or $\bar{N}_{e}^{c}$) 
for a given neutrino trajectory
determined using the PREM (or the Stacey) model,  
reproduces with a remarkably high precision 
the corresponding probability, calculated 
by solving numerically the 
relevant system of evolution equations
with the much more sophisticated Earth density profile
of the PREM (or Stacey) model. 
  
    In the two-layer model, the oscillations of atmospheric 
neutrinos crossing only the Earth mantle (but not 
the Earth core), correspond to oscillations in matter with
constant density. The relevant expressions for 
$P_{2\nu}$, $\kappa$ and 
$A_{2\nu}(\nu_{\tau} \rightarrow \nu_{\tau})$
are given by (see, e.g., ref.~\cite{JBSP203}):
%%%%%%%%%%%%%%%%%%%%%%%%
\begin{equation}
P_{2\nu}(\deltaatm, \theta_{13};E, \theta_{n})
=  \sin^2{\left(\frac{\Delta M^2 L}{4 E}\right)} \sin^2 2\theta'_{m}, 
\label{p2n}
\end{equation}
%%%%%%%%%%%%%%%%%%%%%%%%
%%%%%%%%%%%%%%%%%%%%%%%%
\begin{equation}
\kappa \cong {1\over {2}} [ {\deltaatm\over{2 E}}~ L +
\sqrt{2} G_F \bar{N}_e^{man} L - \frac{\Delta M^2 L}{2 E}],
\label{kappa}
\end{equation}
%%%%%%%%%%%%%%%%%%%%%%%%%
%%%%%%%%%%%%%%%%%%%%%%%%%
\begin{equation}
A_{2\nu}(\nu_{\tau} \rightarrow \nu_{\tau}) = 1 +~
\left( e^{-i \frac{\Delta M^2 L}{2 E}} - 1 \right) 
\cos^2\theta'_{m}~
\label{A2nu}
\end{equation}
%%%%%%%%%%%%%%%%%%%%%%%%%%%%
\noindent Here
%%%%%%%%%%%%%%%%%%%%%%%%%%%
\begin{equation}
\Delta M^2 = \\ \nonumber
\deltaatm~
\sqrt{\left( 1 - \frac{\bar{\rho}_{man~}}{\rho^{res}_{man~}}
\right)^2 \cos^22\theta_{13} 
+ \sin^22\theta_{13} }~,~
\label{DE}
\end{equation}
%%%%%%%%%%%%%%%%%%%%%%%
\noindent is the mass difference between the 
two mass-eigenstate neutrinos in the mantle,
 $\theta'_{m}$ is the mixing angle in the mantle,
%%%%%%%%%%%%%%%%%%
\begin{equation}
\sin^22\theta'_{m} = \frac{\sin^22\theta_{13}}
{(1 - \frac{\bar{\rho}_{man}}{\rho^{res}_{man}})^2 
\cos^22\theta_{13} + 
\sin^22\theta_{13} },
\label{thetam}
\end{equation}
%%%%%%%%%%%%%%%%%%%%%
\noindent $L$ is the distance the neutrino
travels in the mantle, $\bar{\rho}_{man}$
and $\rho^{res}_{man}$
are the mean density along 
the neutrino trajectory 
and the resonance density
in the mantle, 
%%%%%%%%%%%%%%%%%%%%%%%%%
\begin{equation}
\rho^{res}_{man} = \frac{\deltaatm \cos2\theta_{13}}{2E\sqrt{2}
G_F Y_e^{man}}~m_{N}~.
\label{rhores}
\end{equation}
%%%%%%%%%%%%%%%%%%%%%%%%
\noindent 
For a neutrino trajectory which is 
specified by a given Nadir angle $\theta_n$ we have:
%%%%%%%%%%%%%%%%%%
\begin{equation}
L = 2 R_{\oplus}\cos \theta_n 
\end{equation}
%%%%%%%%%%%%%%%%%%%
\noindent where $R_{\oplus} = 6371~$km is 
the Earth radius (in the PREM~\cite{PREM81} and 
Stacey~\cite{Stacey:1977} models)
\footnote{Neutrinos cross only the Earth mantle 
on the way to the detector if 
$\theta_{n} \gtap 33.17^{o}$.}.  

   Consider for definiteness the case of $\deltaatm > 0$.
It follows from eqs.~(\ref{Phie}) and~(\ref{Phimu})
that the oscillation effects of interest 
will be maximal if $P_{2\nu} \cong 1$.
The latter is possible provided 
i) the well-known 
resonance condition~\cite{BPPW80,MS85}, leading  
to $\sin^22\theta_m \cong 1$, is fulfilled, and  
ii) $\cos{\left(\frac{\Delta M^2 L}{2 E}\right)} \cong -1$.
Given the values of  
$\bar{\rho}_{man}$ and $Y_e^{man}$, 
or the value of $\bar{N}_e^{man}$,
the first condition determines the
neutrino energy at which 
$P_{2\nu}$ can be enhanced:
%%%%%%%%%%%%%%%%%%%%%%%%%%%%%%
\begin{equation}
E_{res} \cong 6.6 \left(\frac{\deltaatm}{10^{-3}~{\rm eV^2}}\right)
\left(\frac{\rm {N_A cm^{-3}}}{\rm \bar{N}_e^{man}}\right)
  \cos2\theta_{13}~{\rm GeV}~.
\label{Eres}
\end{equation}
%%%%%%%%%%%%%%%%%%%%%%%%%%%%
\noindent 
If the first condition is satisfied, the second 
determines the length of the path of the neutrinos 
in the mantle for which one can have
$P_{2\nu} \cong 1$:
%%%%%%%%%%%%%%%%%%%%%%%%%%%%%%
\begin{equation}
\left(\frac{\Delta M^2 L}{2 E}\right)_{res} \cong
 1.2\pi~\tan2\theta_{13}~   
 \left(\frac{\rm{\bar{N}_e^{man}}}{\rm{N_A cm^{-3}}}\right)
  \left(\frac{L}{10^{4}~{\rm km}}\right) = \pi , 
\label{Xman}
\end{equation}
%%%%%%%%%%%%%%%%%%%%%%%%%%%%
\noindent 
Taking $\deltaatm \cong (2.0 - 3.0) \times 10^{-3}~{\rm eV^2}$,
${\rm \bar{N}_e^{man} \cong 2~{\rm N_A cm^{-3}}}$
and $\cos2\theta_{13} \cong 1$ one finds 
from eq.~(\ref{Eres}): $E_{res} \cong (6.6 - 10.0)~{\rm GeV}$.
The width of the resonance in $E$, 2$\delta E$,
is determined, as is well-known, 
by $\tan2\theta_{13}$:
$\delta E/E_{res} \sim \tan 2\theta_{13}$. 
For $\sin^2\theta_{13} \sim (0.01 - 0.05)$,
the resonance is relatively wide in the neutrino energy:
$\delta E/E_{res} \cong (0.27 - 0.40)$.
Equation~(\ref{Xman}) implies that 
for $\sin^2\theta_{13} = 0.05~(0.025)$ 
and $\bar{N}_e^{man} \cong 2.2~{\rm N_A cm^{-3}}$,
one can have $P_{2\nu} \cong 1$
only if $L \cong 8000~(10000)~{\rm km}$.

   It follows from the above simple analysis~\cite{mantle} 
that the Earth matter effects can  
amplify $P_{2\nu}$
significantly when the neutrinos cross 
only the mantle i) for 
$E \sim (6 - 11)$ GeV, i.e., in the
multi-GeV range of neutrino energies,
and ii) only for sufficiently long 
neutrino paths in the mantle, i.e., for
$\cos\theta_n \gtap 0.3$.
The magnitude of the matter effects 
of interest increases with increasing 
of $\sin^2\theta_{13}$.

  In the case of atmospheric 
neutrinos crossing the Earth core, new
resonant effects become apparent. 
For $\sin^2\theta_{13} < 0.05$ and
$\deltaatm > 0$,
 we can have $P_{2\nu} \cong 1$ 
{\it only due to the effect of maximal constructive 
interference between the amplitudes of the 
the $\nu_{e} \rightarrow \nu'_{\tau}$
transitions in the Earth mantle and in the 
Earth core}~\cite{SP3198,106,107}.
The effect differs from the MSW one~\cite{SP3198} and the 
enhancement happens in the case of interest at 
a value of the energy  between the resonance energies 
corresponding to the density in the mantle 
and that of the core.
The {\it mantle-core enhancement effect} 
is caused by the existence 
(for a given neutrino trajectory
through the Earth core) of 
{\it points of resonance-like 
total neutrino conversion}, 
$P_{2\nu} = 1$,
in the corresponding space 
of neutrino oscillation 
parameters~\cite{106,107}. 
The points where $P_{2\nu} = 1$
are determined by the conditions~\cite{106,107}:
%%%%%%%%%%%%%%%%%%%%%%%%%%%%
\begin{equation}
\tan \phi'
\pm \sqrt{\frac{-\cos2\theta_m''}{\cos(2\theta_m'' - 4\theta_m')}},~~
\tan \phi''= \pm \frac{\cos2\theta_m'}
{\sqrt{-\cos2\theta_m''\cos(2\theta_m'' - 4\theta_m')}},~~
\label{NOLR1}
\end{equation}
%%%%%%%%%%%%%%%%%%%%%%%%%%%
\noindent where the signs are correlated and 
$\cos2\theta_m''\cos(2\theta_m'' - 4\theta_m') \leq 0$.
In eq.~(\ref{NOLR1}) $2\phi'$ and $2\phi''$
are the oscillation phases
(phase differences) accumulated 
by the (two) neutrino states
after crossing respectively the 
first mantle layer and
the core, and $\theta_m''$ is 
the neutrino mixing angle in the core. 
A rather complete set of values of 
$\deltaatm/E$ and $\sin^22\theta_{13}$
for which both conditions in eq.~(\ref{NOLR1}) 
hold and $P_{2\nu} = 1$
for the Earth core-crossing atmospheric 
$\nu_{\mu}$ and $\nu_{e}$
having trajectories 
with Nadir angle $\theta_n = 0;~13^0;~23^0;~30^0$
was found in ref.~\cite{107}. 
The location of these points 
determines the regions
where $P_{2\nu}$ 
is large, $P_{2\nu} \gtap 0.5$. 
These regions vary slowly with the Nadir angle,
they are remarkably wide in the Nadir angle
and are rather wide in the neutrino energy~\cite{107},
so that the transitions of interest produce noticeable 
effects: we have $\delta E/E \cong 0.3$  
for the values of $\sin^2\theta_{13}$
of interest~\cite{SPNu98,107}.

  For $\sin^2\theta_{13} < 0.05$,
there are two sets of values of 
$\deltaatm$ and $\sin^2\theta_{13}$
for which eq.~(\ref{NOLR1})
is fulfilled and $P_{2\nu} = 1$.
These two solutions of eq.~(\ref{NOLR1})
occur for, e.g., $\theta_n = 0;~13^0;23^0$,
at  1) $\sin^22\theta_{13} = 0.034;~0.039;~0.051$, 
$\deltaatm/E = 7.2;~7.0;~6.5~\times10^{-7}~{\rm eV^2/MeV}$,
and at 2) $\sin^22\theta_{13} = 0.15;~0.17;~0.22$, 
$\deltaatm/E = 4.8;~4.5;~3.8~ \times 10^{-7}~{\rm eV^2/MeV}$
(see Table 2 in ref. \cite{107}).
The first solution corresponds to~\cite{SP3198}
$\cos 2\phi' \cong -1$,
$\cos 2\phi''\cong -1$ and
\footnote{The term
``neutrino oscillation length resonance'' (NOLR) 
was used in ref.~\cite{SP3198} to denote the mantle-core 
enhancement effect in this case.}
$\sin^2(2\theta_m'' - 4\theta_m') = 1$. 
For $\deltaatm = 2.0~(3.0)\times 10^{-3}~{\rm eV^2}$, 
the total neutrino conversion
occurs in the case of the first solution
at $E \cong (2.8 - 3.1)~{\rm GeV}$
($E \cong (4.2 - 4.7)~{\rm GeV}$).
The values of $\sin^22\theta_{13}$ 
at which the second
solution takes place are marginally allowed.
If, e.g., $\deltaatm = 2.5\times 10^{-3}~{\rm eV^2}$,
one has $P_{2\nu} = 1$ for this solution 
for a given $\theta_n$  in the interval
$0 \ltap \theta_n \ltap 23^0$
at $E$ lying in the interval
$E \cong (5.3 - 6.7)~{\rm GeV}$.
  
  The effects of the mantle-core 
enhancement of $P_{2\nu}$ 
(or $\bar{P}_{2\nu}$) 
increase rapidly with $\sin^22\theta_{13}$
as long as $\sin^22\theta_{13}\ltap 0.06$,
and should exhibit a rather weak dependence on
$\sin^22\theta_{13}$ for
$0.06 \ltap \sin^22\theta_{13} < 0.19$.
If 3-neutrino oscillations of 
atmospheric neutrinos take place,
the magnitude of the matter effects 
in the multi-GeV $\mu-$like and $e-$like 
event samples, produced by neutrinos 
crossing the Earth core, 
should be larger than 
in the event samples
due to neutrinos
crossing only the Earth mantle 
(but not the core).
This is a consequence of the fact 
that in the energy range of interest
the atmospheric neutrino fluxes 
decrease rather rapidly with
energy - approximately as $E^{-2.7}$,
while the neutrino interaction 
cross section rises only linearly 
with $E$, and that the maximum
of $P_{2\nu}$ (or $\bar{P}_{2\nu}$)
due to the resonance-like 
mantle-core interference effect
takes place at approximately 
two times smaller energies 
than that due to the MSW effect 
for neutrinos crossing only the
Earth mantle (e.g., at 
$E \cong (3.5 - 3.9)~{\rm GeV}$
and $E \cong 8.3~{\rm GeV}$, respectively,
for $\deltaatm = 2.5\times 10^{-3}~{\rm eV^2}$).

  The same results, eqs.~(\ref{Eres}) and~(\ref{Xman}),
and conclusions are valid for the antineutrino 
oscillation probability $\bar{P}_{2\nu}$ in the case of 
$\deltaatm < 0$.
As a consequence, a preferable detector
for distinguishing the type of mass hierarchy
would be the one with muon charge discrimination, 
such that neutrino interactions can be 
distinguished from those due to antineutrinos. 

%%%%%%%%%%%%%%%%%%%%%%%%%%%%%%%%%
\section{Results}
%%%%%%%%%%%%%%%%%%%%%%%%%%%%%%%%%
\hskip 0.6truecm  
 It follows from  the preceding analysis that
in the case of detectors with muon charge identification,
as observable which is most sensitive 
to the Earth matter effects, and thus to the value
of $\sin^2\theta_{13}$ and the sign of 
$\deltaatm$, as well as to $\sin^2\theta_{23}$, 
we can consider the
Nadir-angle ($\theta_n$) distribution of the 
ratio $N(\mu^{-})/N(\mu^+)$ 
of the multi-GeV $\mu^-$ and  $\mu^+$ 
event rates, or equivalently the 
Nadir-angle distribution
of the $\mu^- - \mu^+$ 
event rate asymmetry
%%%%%%%%%%%%%%%%%%%%%%%%%%
\begin{equation}
A_{\mu^-\mu^+} = \frac{N(\mu^{-}) - N(\mu^+)}{N(\mu^{-}) + N(\mu^+)}~.
\label{Asym}
\end{equation}
%%%%%%%%%%%%%%%%%%%%%%%%%%%%%%%
\noindent We have obtained predictions for the 
$\cos\theta_n$ distribution of 
the ratio $N(\mu^{-})/N(\mu^+)$ and 
the asymmetry $A_{\mu^-\mu^+}$
in the case of 3-neutrino
oscillations of the atmospheric 
$\nu_{\mu}$, $\bar{\nu}_{\mu}$,
$\nu_e$ and $\bar{\nu}_e$,
both for neutrino mass spectra with 
normal ($\deltaatm > 0$)
and inverted  ($\deltaatm < 0$) hierarchy,
and for $\sin^2\theta_{23} = 0.64;~0.50;~0.36$,
and $\sin^22\theta_{13} = 0.05;~0.10$.
These are compared  with the predicted
Nadir-angle distributions 
of the same ratio and 
asymmetry in the case of 2-neutrino 
($\sin^2\theta_{13} = 0$) vacuum
$\nu_{\mu} \rightarrow \nu_{\tau}$ and
$\bar{\nu}_{\mu} \rightarrow \bar{\nu}_{\tau}$
oscillations  of the atmospheric $\nu_{\mu}$ and 
$\bar{\nu}_{\mu}$,
$A^{2\nu}_{\mu^-\mu^+}$.

\begin{figure}
\includegraphics[height=18.2cm]{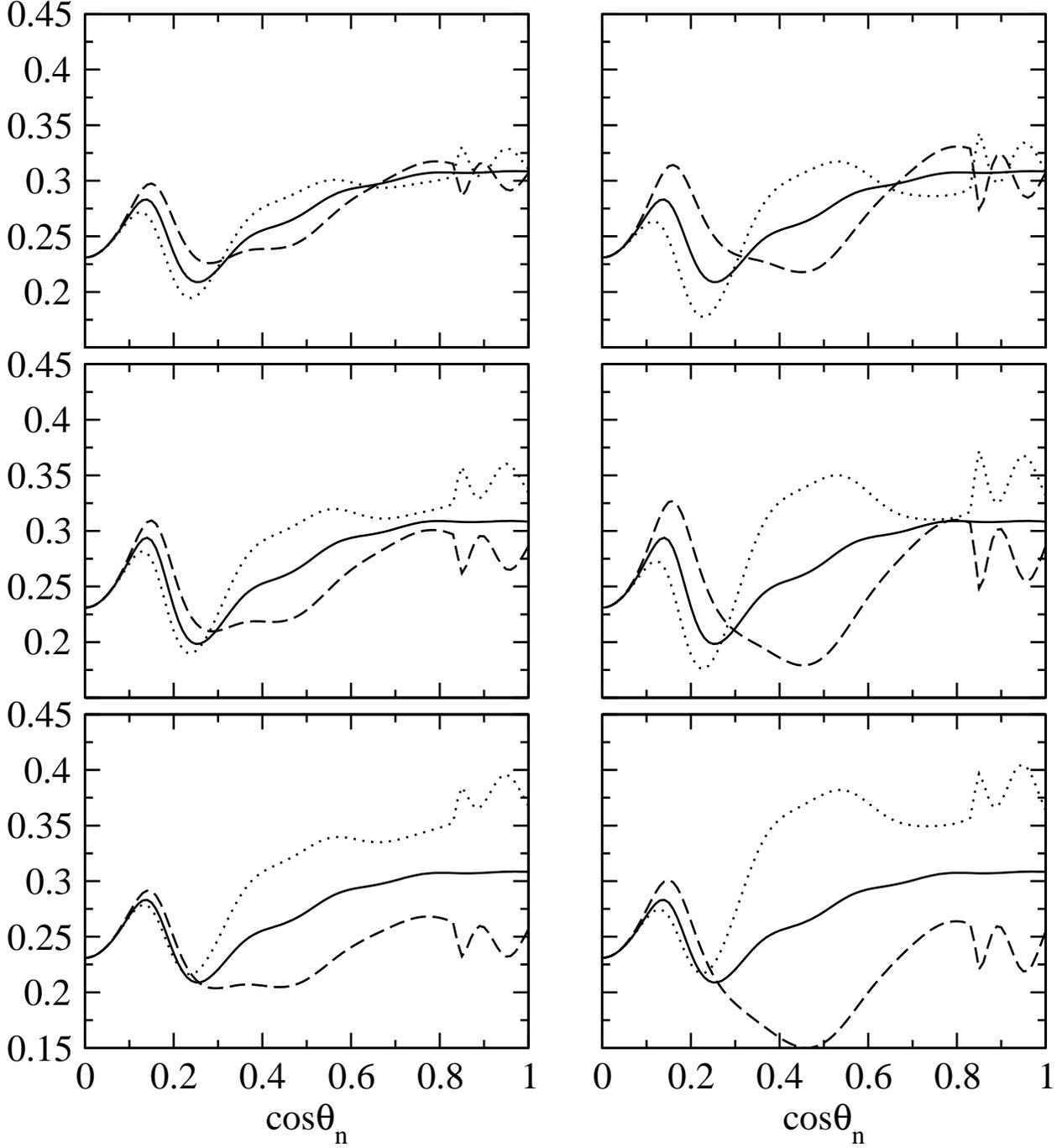}
\caption{\label{dist2210} The Nadir angle distribution
of the charge asymmetry,
$A_{\mu^-\mu^+}$, eq.~(\ref{Asym}),
of the multi-GeV
$\mu^-$ and $\mu^+$ event rates, integrated
over the neutrino (and muon) energy 
in the interval $E = (2.0 - 10.0)$ GeV, 
in the cases 
i) of 2-neutrino
$\nu_{\mu} \rightarrow \nu_{\tau}$ 
and $\bar{\nu}_{\mu} \rightarrow \bar{\nu}_{\tau}$
oscillations in vacuum 
of the atmospheric $\nu_{\mu}$ and
$\bar{\nu}_{\mu}$  and
no $\nu_e$ and $\bar{\nu}_e$ oscillations,
$A_{\mu^-\mu^+}^{2\nu}$ (solid lines),
ii) 3-neutrino
oscillations of 
$\nu_{\mu}$, $\bar{\nu}_{\mu}$
$\nu_e$ and $\bar{\nu}_e$ in the Earth
and neutrino mass spectrum
with normal hierarchy 
$(A_{\mu^-\mu^+}^{3\nu})_{\rm NH}$ (dashed lines),
or with inverted hierarchy,
$(A_{\mu^-\mu^+}^{3\nu})_{\rm IH}$ (dotted lines).
The results shown are
for $|\deltaatm| = 2\times 10^{-3}~{\rm eV^2}$,
$\sin^2\theta_{23} = 0.36~(\rm upper~panels);~0.50
~(\rm middle~panels);~0.64~(\rm lower~panels)$,
and $\sin^22\theta_{13} = 0.05~(\rm left~panels);
~0.10~(\rm right~panels)$.
}
\end{figure}

\begin{figure}
\includegraphics[height=18.2cm]{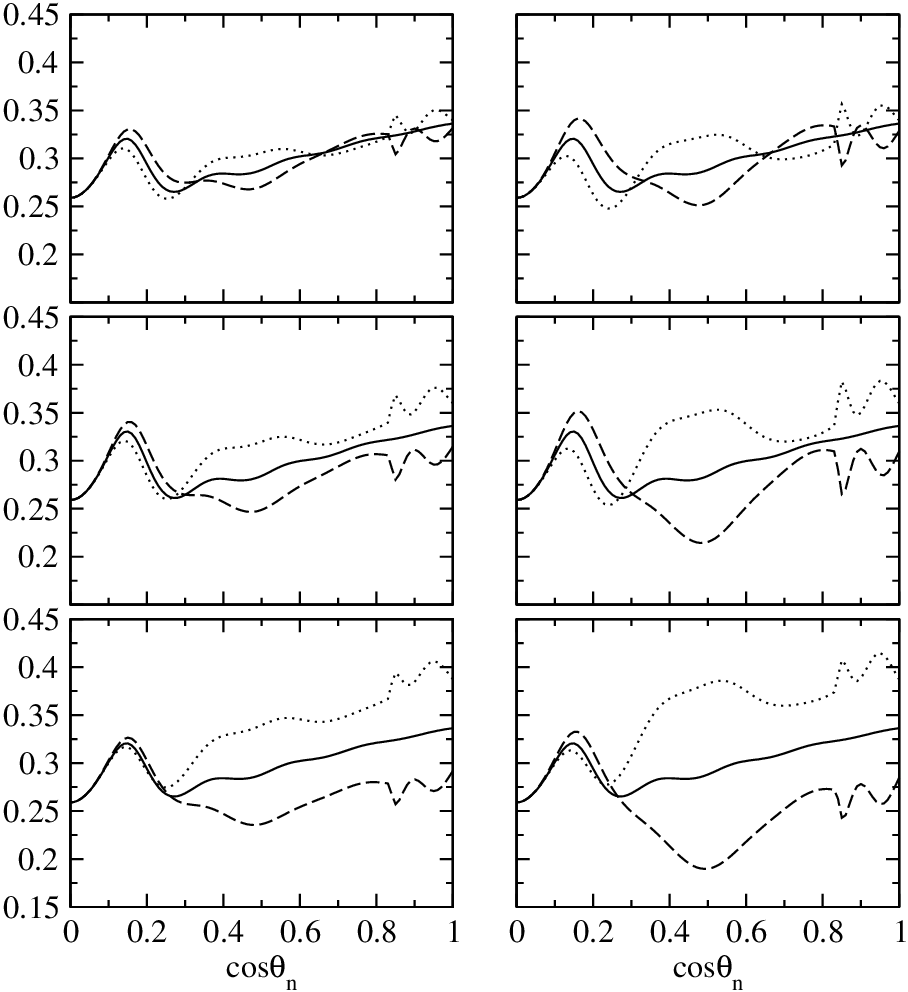}
\caption{\label{dist2220} The same as in Fig.~\ref{dist2210}, 
but for $\mu^-$ and $\mu^+$ event rates integrated
over the neutrino (and muon) energy 
in the interval $E = (2.0 - 20.0)$ GeV.   
}
\end{figure}

\begin{figure}
\includegraphics[height=18.2cm]{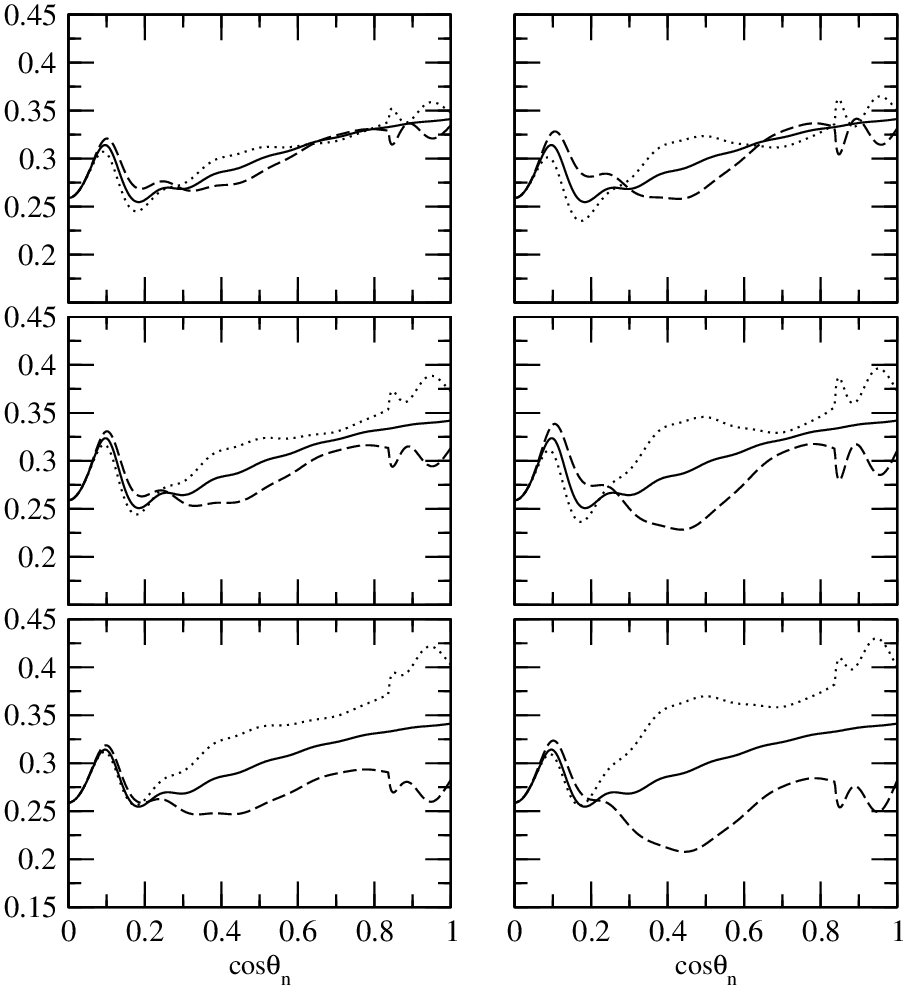}
\caption{\label{dist3220} The same as in Fig.~\ref{dist2210}, but 
for $\mu^-$ and $\mu^+$ event rates integrated
over the neutrino (and muon) energy in the interval $E = (2.0 - 20.0)$
GeV and $|\deltaatm| = 3\times 10^{-3}~{\rm eV^2}$. 
}
\end{figure}

\begin{figure}
\includegraphics[height=18.2cm]{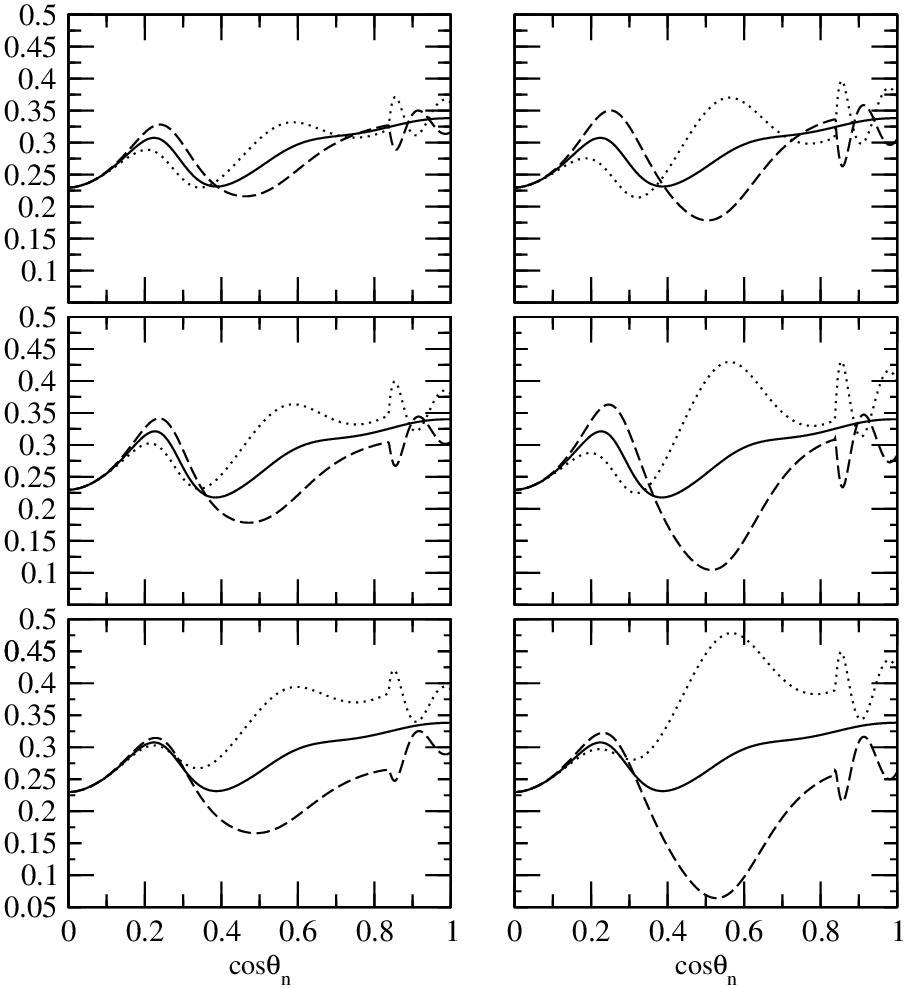}
\caption{\label{dist3520} The same as in Fig.~\ref{dist2210}, but 
for $\mu^-$ and $\mu^+$ event rates integrated
over the neutrino (and muon) energy in the interval $E = (5.0 - 20.0)$
GeV and for $|\deltaatm| = 3\times 10^{-3}~{\rm eV^2}$.
}
\end{figure}

\begin{figure}
\includegraphics[height=18.2cm]{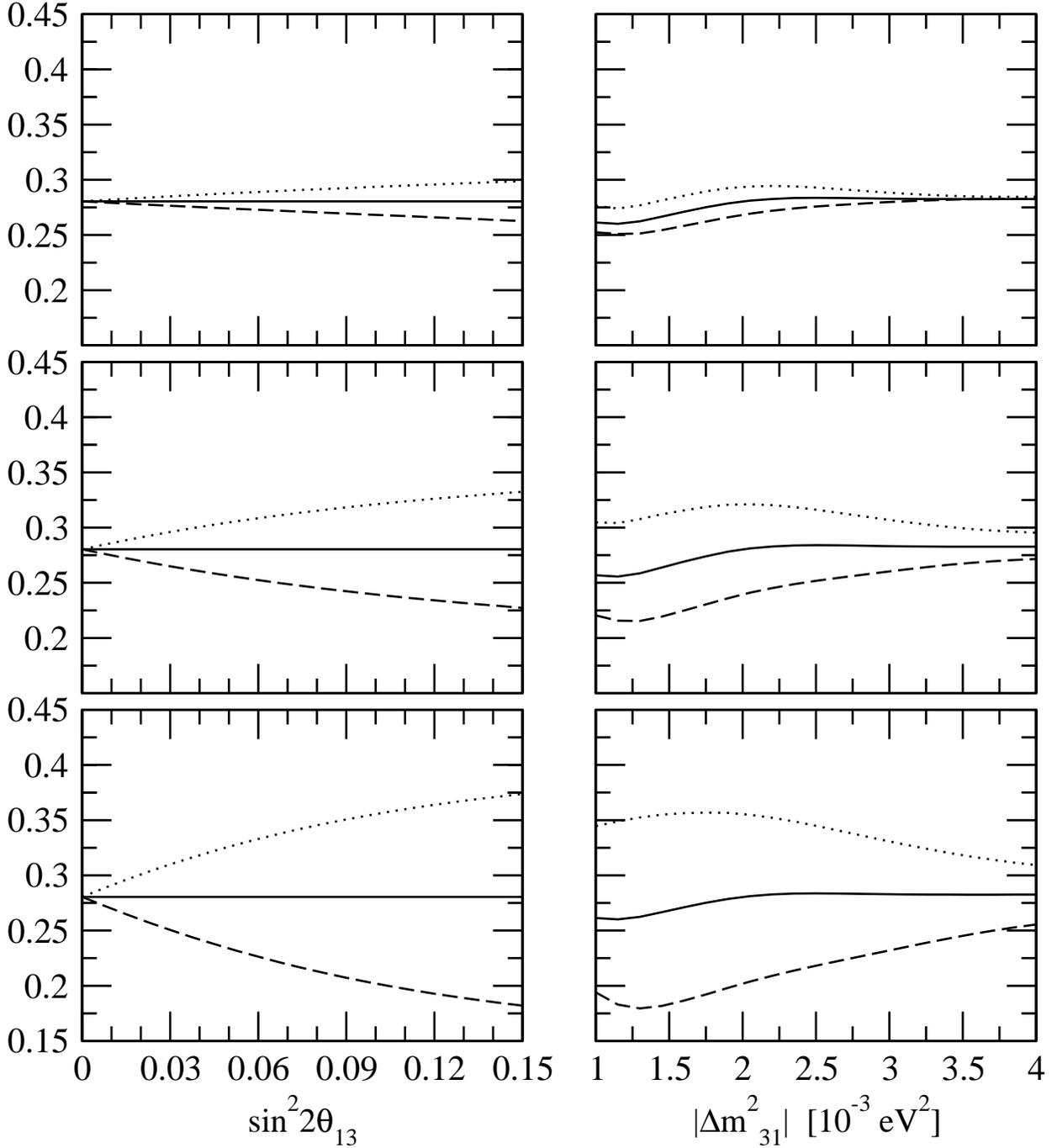}
\caption{\label{integ2210man} The charge asymmetry
$A_{\mu^-\mu^+}$ 
of the multi-GeV
$\mu^-$ and $\mu^+$ event rates,
integrated over the neutrino (and muon) energy 
in the interval $E = (2.0 - 10.0)$ GeV and over
the Nadir angle in the interval corresponding to
$0.30 \leq \cos{\theta_n} \leq 0.84$ ({\it mantle bin}),
as function i) of $\sin^22\theta_{13}$ 
for $|\deltaatm| = 2\times 10^{-3}~{\rm eV^2}$ 
(left panels), and ii) of 
$|\deltaatm|$ for 
$\sin^22\theta_{13} = 0.10$ (right panels):
$A_{\mu^-\mu^+}^{2\nu}$ (solid lines),
$(A_{\mu^-\mu^+}^{3\nu})_{\rm NH}$ (dashed lines) and
$(A_{\mu^-\mu^+}^{3\nu})_{\rm IH}$ (dotted lines).
The results shown are obtained
for $\sin^2\theta_{23} = 0.36~(\rm upper~panels);~0.50
~(\rm middle~panels);~0.64~(\rm lower~panels)$.
}
\end{figure}

\begin{figure}
\includegraphics[height=18.2cm]{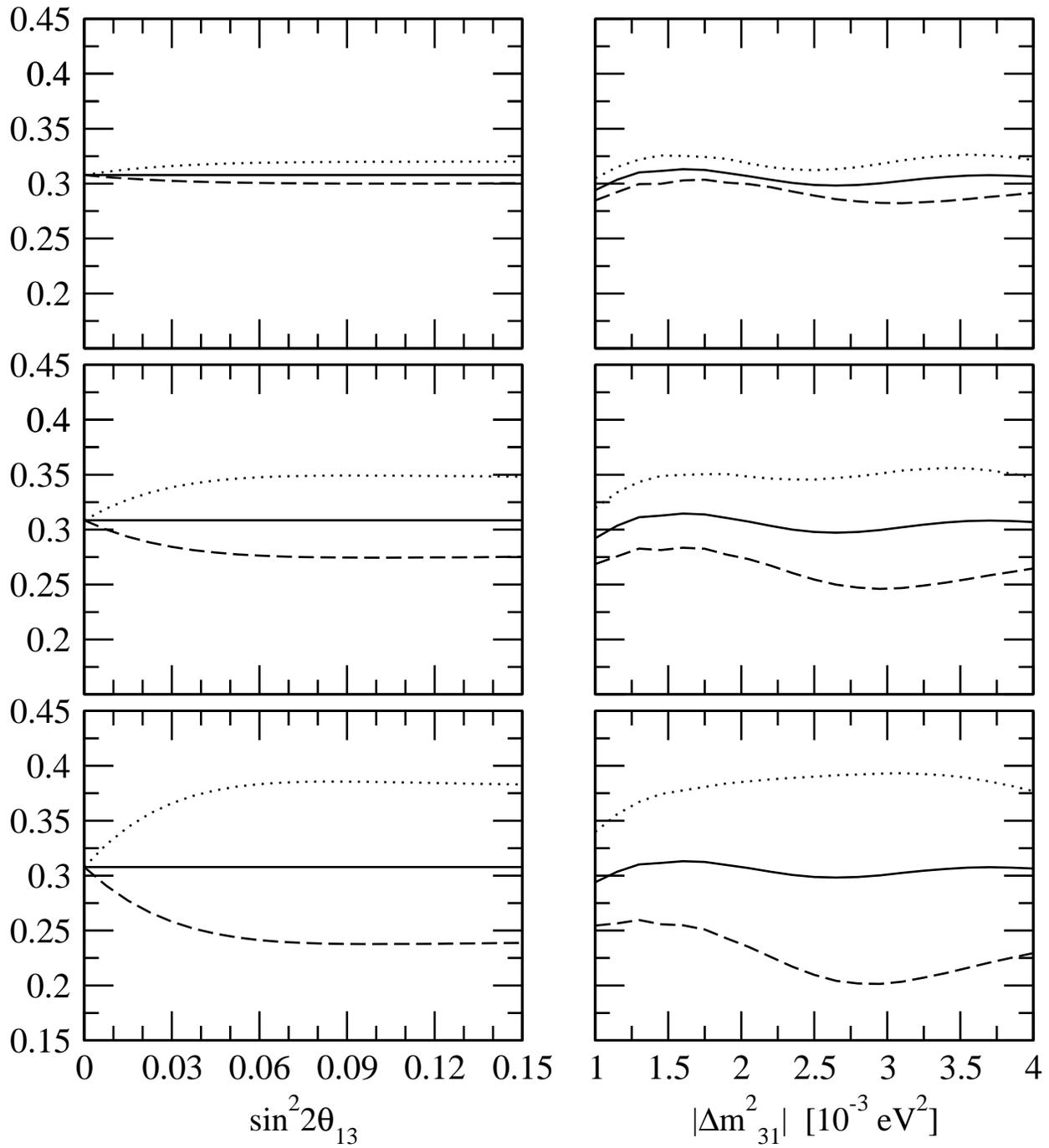}
\caption{\label{integ2210core} The same as in Fig.~\ref{integ2210man},
  but for $\mu^-$ and $\mu^+$ event rates
integrated over the Nadir angle in the interval corresponding to
$0.84 \leq \cos{\theta_n} \leq 1.00$ ({\it core bin}).
}
\end{figure}

\begin{figure}
\includegraphics[height=18.2cm]{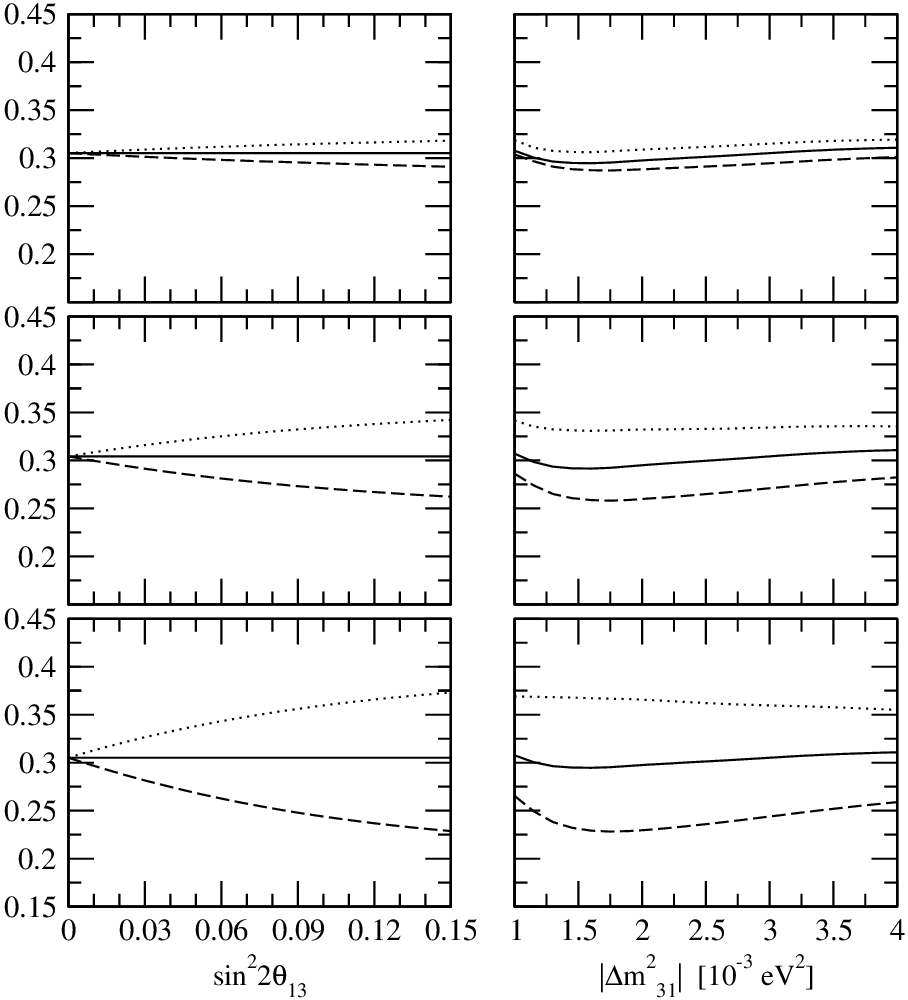}
\caption{\label{integ3220man} The same as in Fig.~\ref{integ2210man},
  but for $|\deltaatm| = 3\times 10^{-3}~{\rm eV^2}$ and 
$\mu^-$ and $\mu^+$ event rates,
integrated over the neutrino (and muon) energy 
in the interval $E = (2.0 - 20.0)$ GeV.
}
\end{figure}

\begin{figure}
\includegraphics[height=18.2cm]{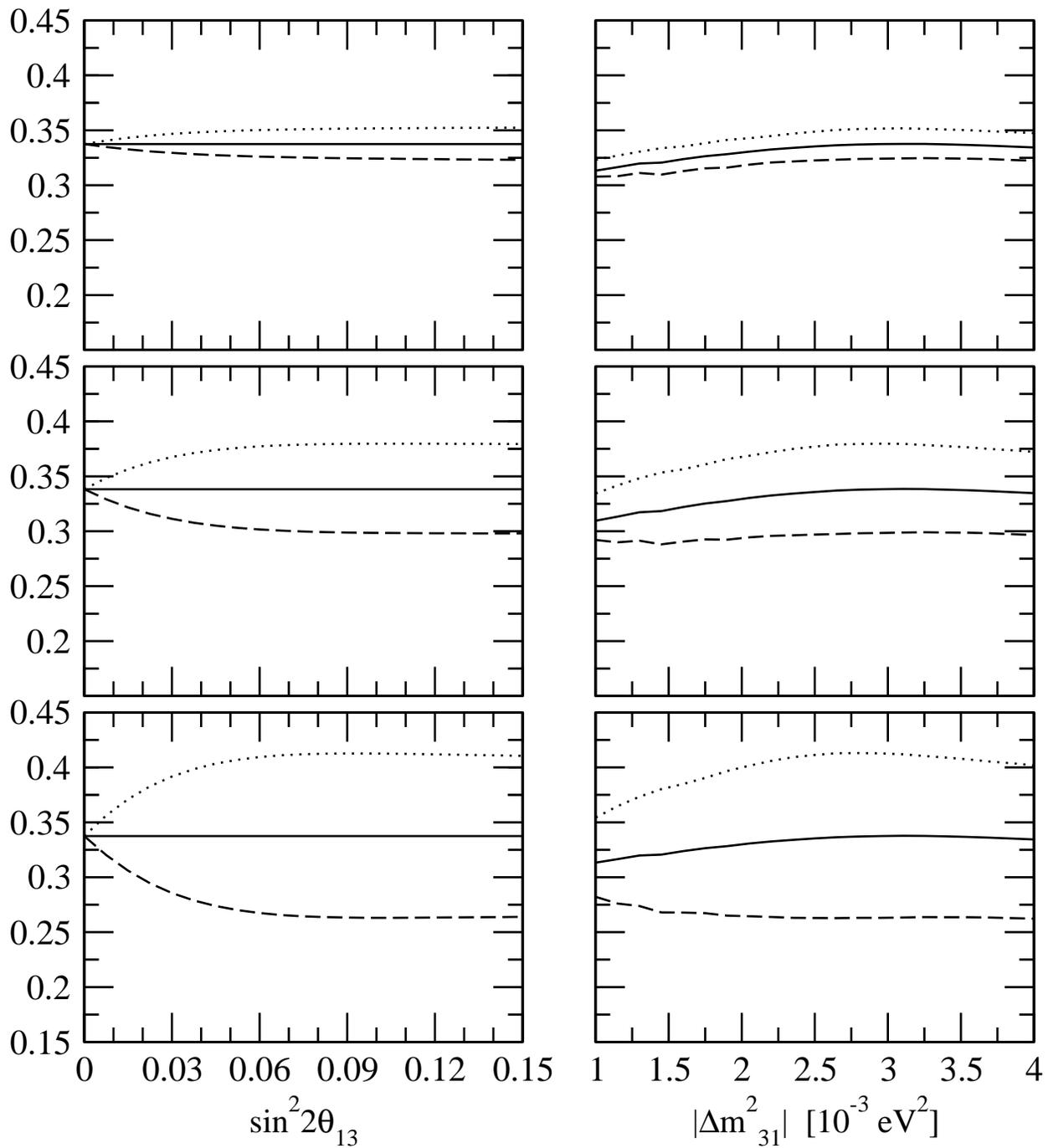}
\caption{\label{integ3220core}  The same as in 
Fig.~\ref{integ2210man}, but 
for $|\deltaatm| = 3\times 10^{-3}~{\rm eV^2}$ and 
$\mu^-$ and $\mu^+$ event rates,
integrated over the neutrino (and muon) energy 
in the interval $E = (2.0 - 20.0)$ GeV and over
the Nadir angle in the interval corresponding to
$0.84 \leq \cos{\theta_n} \leq 1.00$ ({\it core bin}). 
}
\end{figure}

\begin{figure}
\includegraphics[height=18.2cm]{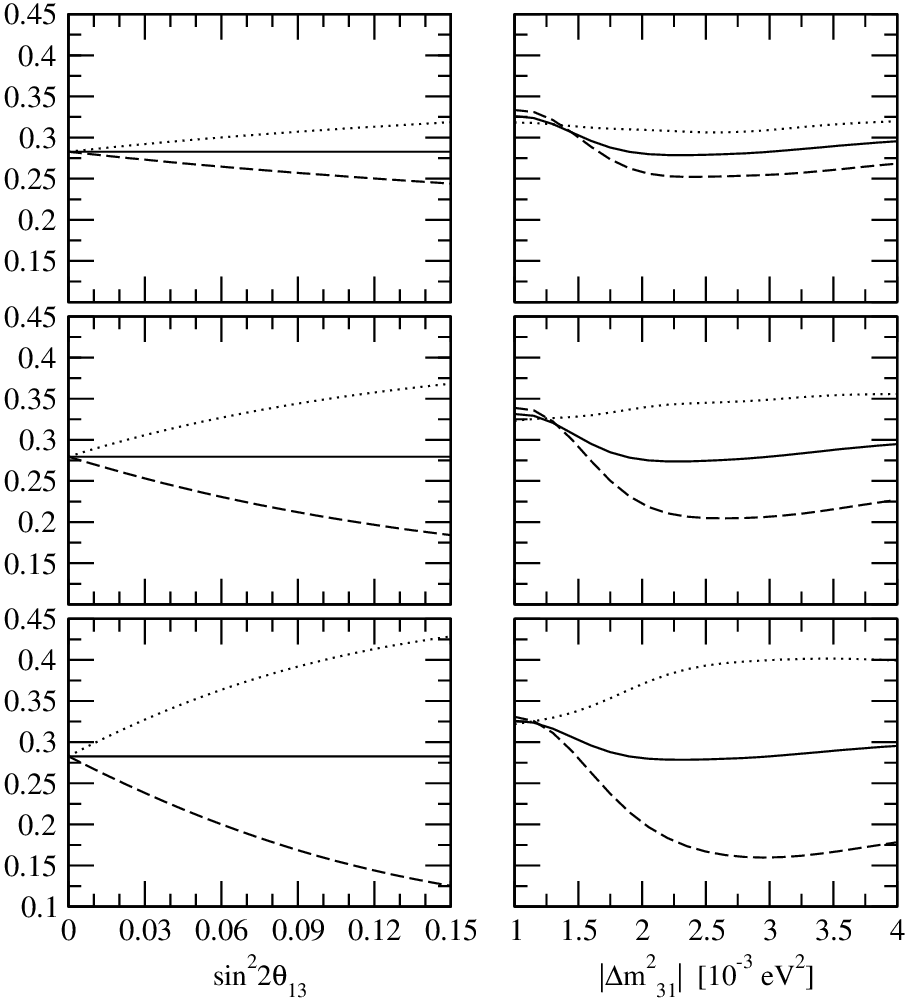}
\caption{\label{integ3520man} The same as in 
Fig.~\ref{integ2210man}, but 
for $\mu^-$ and $\mu^+$ event rates
integrated over the neutrino (and muon) energy 
in the interval $E = (5.0 - 20.0)$ GeV,
 and for $|\deltaatm| = 3\times  10^{-3}~{\rm eV^2}$. 
}
\end{figure}

  In the calculations we have used
the predictions for the Nadir angle and 
energy distributions of the
atmospheric neutrino fluxes 
given in ref.~\cite{Naumov}. 
The interactions of the atmospheric 
neutrinos are described 
by taking into account 
only the $\nu_{\mu}$ and $\bar{\nu}_{\mu}$ 
deep inelastic scattering (DIS) 
cross sections. The latter are calculated
using the GRV94 parton distributions 
given in ref.~\cite{GRV94}. 
We present here results for the asymmetry
$A_{\mu^-\mu^+}$~~
\footnote{It is interesting to note
that the ratio $N(\mu^{-})/N(\mu^+)$ exhibits
essentially the same dependence on 
$\cos\theta_n$ as the asymmetry
$A_{\mu^-\mu^+}$. This is a consequence of
the special form of the dependence
of $A_{\mu^-\mu^+}$ on $N(\mu^{-})/N(\mu^+)$
and of the fact that typically one finds
$N(\mu^{-})/N(\mu^+) \sim (1.5 - 2.4)$ 
for the ranges of the values of 
the parameters of interest.
Correspondingly, the following 
approximate relation holds (within
$\sim 20\%$ and typically 
with much higher precision)
for the range
of values of the parameters of interest: 
$N(\mu^{-})/N(\mu^+) \cong
6~A_{\mu^-\mu^+}$.}. 
They are shown graphically
in Figs.~\ref{dist2210} -~\ref{integ3520man}. 
The figures correspond 
to three different intervals of 
integration over the energies of the 
atmospheric $\nu_{\mu}$ and $\bar{\nu}_{\mu}$,
and of the $\mu^-$ and $\mu^+$ they produce 
in the detector,
$E = [2,10],~[2,20],~[5,20]$ GeV, and 
\footnote{The iron-magnetized calorimeter
detectors allow to reconstruct
with a certain precision
the initial neutrino energy as well,
see refs.~\cite{MINOS,INO,MONOLITH}.}
thus to three different 
possible event samples.      
Figures~\ref{dist2210} -~\ref{dist3520} show the 
the asymmetries $A_{\mu^-\mu^+}$ and $A^{2\nu}_{\mu^-\mu^+}$
as functions of
$\cos\theta_n$ for two ``reference'' values of 
$|\deltaatm|$, $\deltaatm = \pm 2\times 10^{-3}~{\rm eV^2}$
and $\deltaatm = \pm 3\times 10^{-3}~{\rm eV^2}$,
while in Figs.~\ref{integ2210man} -~\ref{integ3520man}
we present results for
the asymmetries in the rates of 
the multi-GeV $\mu^-$ and  $\mu^+$ 
events, integrated over 
$\cos\theta_n$ in the intervals
[0.30,0.84] ({\it mantle bin}) and
[0.84,10] ({\it core bin}),
$\bar{A}_{\mu^-\mu^+}$ and $\bar{A}^{2\nu}_{\mu^-\mu^+}$.
The dependence of the latter 
on $\sin^22\theta_{13}$ for 
$\deltaatm = \pm 2\times 10^{-3}~{\rm eV^2}$, 
and on $\deltaatm$ for 
$\sin^22\theta_{13} = 0.10$,
is shown for three values of 
$\sin^2\theta_{23} = 0.36;~0.50;~0.64$.

   As Figs.~\ref{dist2210} -~\ref{integ3520man} indicate,
the Earth matter effects can produce
a noticeable deviations of 
$A_{\mu^-\mu^+}$ from the 2-neutrino 
vacuum oscillation
asymmetry $A^{2\nu}_{\mu^-\mu^+}$
at  $\cos\theta_{n} \gtap 0.3$.
As a quantitative measure of the magnitude 
of the matter effects one can use the
deviation of the asymmetry $A_{\mu^-\mu^+}$
in the case of 3-neutrino oscillations,
$\sin^22\theta_{13} \neq 0$,
$\sin^22\theta_{13} \gtap 0.04$,
from the asymmetry, $A^{2\nu}_{\mu^-\mu^+}$,
predicted in the case of
2-neutrino oscillations, i.e., 
for $\sin^22\theta_{13} = 0$,
or the relative difference between
the two asymmetries,
%%%%%%%%%%%%%%%%%%%%%%%%%%%%%%%%
\begin{equation}
\Delta~ =~ \frac{A_{\mu^-\mu^+} - A^{2\nu}_{\mu^-\mu^+}}
{A^{2\nu}_{\mu^-\mu^+}}~~.
\label{Delta}
\end{equation}
%%%%%%%%%%%%%%%%%%%%%%%%%%%%%%%%%%
    The magnitude of the matter effects, or the 
relative difference $\Delta$,
depends critically on the value of
$\sin^2\theta_{23}$: $|\Delta|$ increases 
rapidly with the increasing of 
$\sin^2\theta_{23}$. This is clearly
seen in Figs.~\ref{dist2210} -~\ref{integ3520man}.
The matter effects in $A_{\mu^-\mu^+}$ 
are hardly observable for
$\sin^2\theta_{23} \ltap 0.30$.
For $\cos\theta_{n} \leq 0.84$, i.e.,
in the {\it mantle bin},
the asymmetry difference $|\Delta|$ increases
practically linearly with
$\sin^22\theta_{13}$.
In the case of 
$0.84 \leq \cos\theta_{n} \leq 1.0$,
i.e., in the {\it core bin},
$|\Delta|$ increases rapidly with
$\sin^22\theta_{13}$ until the latter
reaches the value of
$\sin^22\theta_{13} \cong 0.06$.
For values of $\sin^22\theta_{13} \cong (0.06 - 0.15)$,
$\Delta$ is essentially independent of 
$\sin^22\theta_{13}$ and is given by 
its value at  $\sin^22\theta_{13} \cong 0.06$
(Figs.~\ref{integ2210man} -~\ref{integ3520man}). 
The magnitude of the asymmetry
difference $\Delta$ depends weakly on
$\deltaatm$ taking values 
in the interval $\sim (2 - 3)\times 10^{-3}~{\rm eV^2}$,
as long as the energy integration interval
is sufficiently wide to include the 
energy regions where the Earth matter
effects enhance strongly the subdominant
transition probabilities. If this is 
not the case, a noticeable
dependence on $\deltaatm$
can be present. This is illustrated 
e.g, in Fig.~\ref{integ2210man}, which 
corresponds to $E = [2,10]$ GeV.
The asymmetry difference in the
{\it mantle bin}
diminishes monotonically as 
$|\deltaatm|$ increases
starting from the value of
 $\sim 1.3\times 10^{-3}~{\rm eV^2}$
and becomes rather small at 
$|\deltaatm|\gtap 3\times 10^{-3}~{\rm eV^2}$.
This behavior can be easily understood:
for $|\deltaatm|\cong 2\times 10^{-3}~{\rm eV^2}$,
the region of enhancement of the subdominant 
neutrino oscillations lies
in the region of energy integration,
while for $|\deltaatm| > 3\times 10^{-3}~{\rm eV^2}$
the enhancement region is
practically outside the region
of integration over the neutrino 
energy.

  For the ranges considered of the three
oscillation parameters,
$\sin^2\theta_{23}$, $\sin^22\theta_{13}$ 
and $|\deltaatm|$,
the magnitude of the asymmetry
difference $|\Delta|$
depends weakly on the
{\it maximal} neutrino (and muon) energy,
$E_{max}$,
for the chosen event sample
as long as $E_{max} \gtap 10$ GeV. 
By increasing the
{\it minimal} energy of the neutrinos
contributing to the event sample,
$E_{min}$,
from 2 GeV to, e.g., 5 GeV,
one could diminish the asymmetry in 
the {\it core bin} substantially (Fig.~\ref{dist3520}). 
In that case, a large fraction of the region 
of enhancement is not included within the interval of integration. 

  For $\sin^2\theta_{23} \gtap 0.50$,
$\sin^22\theta_{13} \gtap 0.06$
and $|\deltaatm| = 
(2 - 3)\times 10^{-3}~{\rm eV^2}$,
the Earth matter effects 
produce an integrated asymmetry difference 
$|\Delta|$ which is bigger than 
approximately $\sim 15\%$,
can reach the values of $(30 - 35)\%$ (Figs.~\ref{integ2210man}
-~\ref{integ3520man}), 
and thus can be sufficiently large to be observable.
As Figs.~\ref{integ2210man} -~\ref{integ3520man} clearly show, 
the sign of
the relative difference of the integrated 
asymmetries
is anticorrelated with the sign 
of $\deltaatm$:
for $\deltaatm > 0$ we have
$A_{\mu^-\mu^+} < A^{2\nu}_{\mu^-\mu^+}$,
while for $\deltaatm < 0$,
the inequality $A_{\mu^-\mu^+} > A^{2\nu}_{\mu^-\mu^+}$
holds 
\footnote{Note that $A^{2\nu}_{\mu^-\mu^+} > 0$.
This is a consequence of the fact that
the $\nu_{\mu}$ DIS cross section
is approximately by a factor 2 bigger
than the $\bar{\nu}_{\mu}$ DIS cross section and that 
the fluxes of atmospheric 
$\nu_{\mu}$ and $\bar{\nu}_{\mu}$,
($\nu_{e}$ and $\bar{\nu}_{e}$) do not differ
considerably.}.
Therefore the measurement of 
$A_{\mu^-\mu^+}$ can provide
a direct information on the 
sign of $\deltaatm$,
i.e., on the neutrino mass hierarchy.

 It follows from Figs.~\ref{dist2210}-~\ref{dist3520} that
the deviations of the asymmetry 
$A_{\mu^-\mu^+}$ from the 2-neutrino oscillation 
one, $A^{2\nu}_{\mu^-\mu^+}$,
are maximal typically
i) in the ``core bin'',
 $\cos\theta_{n} = [0.84 - 1.0]$,
and ii) at $\cos\theta_{n} \sim 0.50$
in the ``mantle bin'',  
$\cos\theta_{n} \leq 0.84$.
For $\deltaatm = 2\times 10^{-3}~{\rm eV^2}$,
$\sin^2\theta_{23} = 0.50~(0.64)$
and $E = [2,10]$ GeV (Fig.~\ref{dist2210}), we have
at $\cos\theta_{n} \sim 0.50$
and $\sin^22\theta_{13} = 0.05$:
$A^{2\nu}_{\mu^-\mu^+} \cong 0.29$,
while the matter effects
lead to $A_{\mu^-\mu^+} \cong ~0.25~(0.24)$.
This corresponds to a 
negative relative
difference between $A(\mu^-\mu^+)$
and $A^{2\nu}(\mu^-\mu^+)$, $\Delta < 0$, 
and $|\Delta| \sim 14\%~(17\%)$. 
For $\sin^22\theta_{13} = 0.10$,
one finds  $\Delta \sim - 28\%~(34\%)$.
The relative difference $\Delta$ 
in the {\it core bin} is also negative
and $|\Delta|$
has similar or larger values.
For $E$ in the interval $E = [5,20]$ GeV (Fig.~\ref{dist3520}),
we get at $\cos\theta_{n} \sim 0.50$
for $\deltaatm = 3\times 10^{-3}~{\rm eV^2}$,
$\sin^22\theta_{13} = 0.10$ and 
$\sin^2\theta_{23} = 0.64$:
$A^{2\nu}_{\mu^-\mu^+} \cong 0.25$,
$A_{\mu^-\mu^+} \cong 0.075$,
and a relative difference between
the two asymmetries $\Delta \cong - 70\%$.

     Our results show that
the Earth matter effects 
in the Nadir-angle distribution of the 
ratio $N(\mu^{-})/N(\mu^+)$ 
of the rates of multi-GeV $\mu^-$ and $\mu^+$ 
events, or equivalently in the 
Nadir-angle distribution
of the $\mu^{-}-\mu^+$ 
event rate asymmetry $A_{\mu^-\mu^+}$, eq.~(\ref{Asym}),
can be sufficiently large to be observable
in the current and planned experiments
with iron magnetized calorimeter 
detectors which have muon charge
identification capabilities (MINOS, INO, etc.). 

%%%%%%%%%%%%%%%%%%%%%%%%%%%%%%%%%%%%%%%
\section{Conclusions}
%%%%%%%%%%%%%%%%%%%%%%%%%%%%%%%%%%%%%%
\hskip 0.6truecm  
   We have studied the possibilities
to obtain information on the values of 
$\sin^2\theta_{13}$ and $\sin^2\theta_{23}$,
and on the sign of
$\deltaatm$
using the data on atmospheric 
neutrinos, which can be obtained in experiments
with detectors able to measure the charge of the muon
produced in the charged current (CC) reaction
by atmospheric $\nu_{\mu}$ or
$\bar{\nu}_{\mu}$ (MINOS, INO, etc).
The indicated oscillation parameters control 
the magnitude of the Earth 
matter effects in the subdominant
oscillations, 
$\nu_{\mu} \rightarrow \nu_e$ 
($\nu_{e} \rightarrow \nu_{\mu}$) and
$\bar{\nu}_{\mu} \rightarrow \bar{\nu}_e$
($\bar{\nu}_{e} \rightarrow \bar{\nu}_{\mu}$),
of the multi-GeV ($E \sim (2 - 10)$ GeV)
atmospheric neutrinos.  
As observable which is most sensitive 
to the Earth matter effects, 
and thus to the value
of $\sin^2\theta_{13}$ and  and the sign of 
$\deltaatm$, as well as to 
$\sin^2\theta_{23}$, 
we have considered 
the Nadir-angle ($\theta_n$) distribution of the 
ratio $N(\mu^{-})/N(\mu^+)$ 
of the multi-GeV $\mu^-$ and  $\mu^+$ 
event rates, and the corresponding 
$\mu^- - \mu^+$ event rate asymmetry
$A_{\mu^-\mu^+}$, eq.~(\ref{Asym}).
The systematic uncertainty, in particular,
in the Nadir angle dependence of 
$N(\mu^{-})/N(\mu^+)$ and of 
the asymmetry $A_{\mu^-\mu^+}$, can 
be smaller than those on the measured
Nadir angle distributions of the rates of
$\mu^-$ and $\mu^+$ events,
$N(\mu^-)$ and $N(\mu^+)$.
We have obtained predictions for the 
$\cos\theta_n$ distribution of 
the asymmetry $A_{\mu^-\mu^+}$
(and of the ratio $N(\mu^{-})/N(\mu^+)$) 
in the case of 3-neutrino 
oscillations of the atmospheric 
$\nu_{\mu}$, $\bar{\nu}_{\mu}$,
$\nu_e$ and $\bar{\nu}_e$,
both for neutrino mass spectra with 
normal ($\deltaatm > 0$)
and inverted  ($\deltaatm < 0$) hierarchy,
and for $\sin^2\theta_{23} = 0.64;~0.50;~0.36$,
and $\sin^22\theta_{13} = 0.05;~0.10$.
These are compared  with the predicted
Nadir-angle distribution
of the same ratio and 
asymmetry in the case of 2-neutrino 
($\sin^2\theta_{13} = 0$)
vacuum
$\nu_{\mu} \rightarrow \nu_{\tau}$ and
$\bar{\nu}_{\mu} \rightarrow \bar{\nu}_{\tau}$
oscillations  of the atmospheric $\nu_{\mu}$ and 
$\bar{\nu}_{\mu}$, $A^{2\nu}_{\mu^-\mu^+}$.

 Our results are summarized in Figs.~\ref{dist2210}
 -~\ref{integ3520man}. Figures~\ref{dist2210} -~\ref{dist3520}
show the
dependence of  the asymmetries $A_{\mu^-\mu^+}$ 
and $A^{2\nu}_{\mu^-\mu^+}$
on $\cos\theta_n$ for two ``reference'' values of 
$|\deltaatm|$,  
$\deltaatm = \pm 2\times 10^{-3}~{\rm eV^2}$
and $\deltaatm = \pm 3\times 10^{-3}~{\rm eV^2}$,
and for three possible ranges 
of energies of the atmospheric neutrinos,
contributing to the event rates of interest,
$E = [2,10],~[2,20],~[5,20]$ GeV.
In Figs.~\ref{integ2210man} -~\ref{integ3520man}
we present results for
the asymmetries in the rates of 
the multi-GeV $\mu^-$ and  $\mu^+$ 
events, integrated over 
$\cos\theta_n$ in the intervals
[0.30,0.84] ({\it mantle bin}) and
[0.84,10] ({\it core bin}).
We find that for  $\sin^2\theta_{23} \gtap 0.50$,
$\sin^22\theta_{13} \gtap 0.06$
and $|\deltaatm| = 
(2 - 3)\times 10^{-3}~{\rm eV^2}$,
the Earth matter effects 
produce a relative difference 
between the {\it integrated} asymmetries
$\bar{A}_{\mu^-\mu^+}$ and $\bar{A}^{2\nu}_{\mu^-\mu^+}$
which is bigger in absolute value than 
approximately $\sim 15\%$,
can reach the values of $(30 - 35)\%$ (Figs.~\ref{integ2210man}
 -~\ref{integ3520man}),
and thus can be sufficiently large to be observable.
As our results show (Figs.~\ref{integ2210man} -~\ref{integ3520man}),
the sign of 
the indicated asymmetry difference,
$(\bar{A}_{\mu^-\mu^+} - \bar{A}^{2\nu}_{\mu^-\mu^+})$,
is directly related to the sign 
of $\deltaatm$: for $\deltaatm > 0$
we have $(\bar{A}_{\mu^-\mu^+} - \bar{A}^{2\nu}_{\mu^-\mu^+}) < 0$,
while if $\deltaatm < 0$ then
$(\bar{A}_{\mu^-\mu^+} - \bar{A}^{2\nu}_{\mu^-\mu^+}) > 0$
 Therefore the measurement of the Nadir 
angle dependence of $A_{\mu^-\mu^+}$,
or of the  value of 
$\bar{A}_{\mu^-\mu^+}$ in the {\it mantle}
and/or in the {\it core} bins,
can provide a direct information on the 
sign of $\deltaatm$,
i.e., on the neutrino mass hierarchy.

  To summarize, the studies of the oscillations of the 
multi-GeV atmospheric $\nu_{\mu}$ and    
$\bar{\nu}_{\mu}$ in experiments with detectors 
having good muon charge identification 
capabilities (MINOS, INO, etc.),
can provide fundamental information  on the values 
of $\sin^2\theta_{13}$ and $\sin^2\theta_{23}$, and 
on the sign of $\deltaatm$, i.e.,
on the neutrino mass hierarchy.
  
%%%%%%%%%%%%%%%%%%%%%%%%%%%%%%%%%%%%%%%%%%%%%%
\section*{Acknowledgments}
%%%%%%%%%%%%%%%%%%%%%%%%%%%%%%%%%%%%%%%%%%%%%%

We are indebted to  J. Bernab\'eu, T. Kajita, 
A. Mann and S. Wojcicki for useful discussions.
S.T.P. would like to thank  
Prof. T. Kugo, Prof. M. Nojiri and 
the other members of the Yukawa Institute 
for Theoretical Physics (YITP), Kyoto, Japan,
where part of the work on this article was done,
for the kind hospitality extended to him. S.P.-R.
would like to thank the Theory Division at CERN for 
hospitality during the final completion of this work.
This work is supported in part 
by the Italian INFN under the programs
``Fisica Astroparticellare'' (S.T.P.)
and by NASA grant NAG5-13399 (S.P.-R.).

\end{document}